\documentstyle[12pt]{article}

 1
 1
 1

\def\ie{{\em i.e.}}
\def\eg{{\em e.g.}}

\def\beq{\begin{equation}}
\def\eeq{\end{equation}}
\def\bdm{\begin{displaymath}}
\def\edm{\end{displaymath}}

\catcode`\@=11 
\def\coeff#1#2{{\textstyle{#1\over #2}}}

\def\vev#1{\left\langle #1\right\rangle}
\def\lsim{\mathrel{\mathpalette\@versim<}}
\def\gsim{\mathrel{\mathpalette\@versim>}}
\def\@versim#1#2{\vcenter{\offinterlineskip
    \ialign{$\m@th#1\hfil##\hfil$\crcr#2\crcr\sim\crcr } }}
\def\etal{{\em et. al.}}
\def\JL{J. L. Lopez}
\def\DVN{D. V. Nanopoulos}
\def\AZ{A. Zichichi}

\def\r#1{$\bf#1$}
\def\rb#1{$\bf\overline{#1}$}

\def\t1{{\tilde 1}}
\def\ov{\overline}
\def\F{\widetilde F}
\def\Fb{\widetilde{\bar F}}

\def\GeV{\,{\rm GeV}}
\def\TeV{\,{\rm TeV}}

\def\to{\rightarrow}
\def\pb{\,{\rm pb}}
\def\ipb{\,{\rm pb}^{-1}}
\def\ifb{\,{\rm fb}^{-1}}
\def\h{{1\over2}}
\def\q{{1\over4}}
\def\tq{{3\over4}}
\def\rt{{1\over\sqrt{2}}}
\def\NPB#1#2#3{Nucl. Phys. B {\bf#1} (19#2) #3}
\def\PLB#1#2#3{Phys. Lett. B {\bf#1} (19#2) #3}

\def\PRD#1#2#3{Phys. Rev. D {\bf#1} (19#2) #3}
\def\PRL#1#2#3{Phys. Rev. Lett. {\bf#1} (19#2) #3}
\def\PRT#1#2#3{Phys. Rep. {\bf#1} (19#2) #3}
\def\MODA#1#2#3{Mod. Phys. Lett. A {\bf#1} (19#2) #3}
\def\IJMP#1#2#3{Int. J. Mod. Phys. A {\bf#1} (19#2) #3}
\def\TAMU#1{Texas A \& M University preprint CTP-TAMU-#1}

\begin{document}
\baselineskip=16pt

\begin{flushright}
\large
{CTP-TAMU-17/94}\\
May 1994\\
\end{flushright}

\begin{center}
\vglue 1cm
{\Large\bf String and String-Inspired Phenomenology
\footnote{\large Lectures delivered at the XXII ITEP International Winter
School of Physics, Moscow, Russia, February 22 -- March 2, 1994}\\}
\vspace{0.2cm}
\vglue 1cm
{\large JORGE L. LOPEZ\\}
\vglue 1cm
{\large\em Center for Theoretical Physics\\
Department of Physics\\
Texas A\&M University\\
College Station, TX 77843--4242, USA\\}

\vglue 2cm
{\large ABSTRACT}
\end{center}
{\rightskip=3pc
 \leftskip=3pc
\noindent
\large
In these lectures I review the progress made over the last few years in the
subject  of string and string-inspired phenomenology. I take a practical
approach, thereby concentrating more on explicit examples rather than on formal
developments. Topics covered include: introduction to string theory, the
free-fermionic formulation and its general features, generic conformal field
theory properties, $SU(5)\times U(1)$ GUT and string model-building,
supersymmetry breaking, the bottom-up approach to string-inspired models,
radiative electroweak symmetry breaking, the determination of the allowed
parameter space of supergravity models and the experimental constraints on
this class of models, and prospects for direct and indirect tests of
string-inspired models.
}
\newpage
\tableofcontents
\newpage
\setcounter{page}{1}
\pagestyle{plain}

\baselineskip=14pt

\section{Introduction}
The quest for a ``Theory of Everything" has captured the imagination of
many physicists over the years. Unfortunately, such an enterprise is by
definition very ambitious and often driven by grand principles rather than
by compelling experimental information. In fact, progress can be made at
the fastest pace when both bright theoretical insights and clear experimental
data are available simultaneously, such as in the development of Quantum
Mechanics in the early part of this century, and more recently in the
elucidation of the theory of strong and electroweak interactions. The larger
theory that is to contain the Standard Model is in the process of being
elaborated at this time, but experimental data are not helping in the
traditional way since they agree with the Standard Model predictions very
well. Nonetheless, physicists believe that some larger theory must exist.

The search for clues as to the nature of this all-encompassing theory has
followed a path towards larger and larger energies, \ie, grand unification,
supersymmetry, supergravity, and superstrings. This line of thought indicates
that the explanation for all observable phenomena is to be found in the theory
of superstrings. Superstrings have so far one undisputable success: they
provide the only known consistent theory of quantum gravity. However, for
physics at low energies, the more interesting aspect of string theory is
its possible explanation of the Standard Model. This aspect of string theory
has developed over the last several years and is the main subject of these
lectures.

\subsection{Why strings?}
Since string theory is so complicated and as yet still only mildly understood,
the motivations for investing a great deal of time exploring it have to be
spelled out in as clear a way as possible.\footnote{I should note that the
more theoretically inclined students seem to find enough motivation to study
string theory in its intriguing and partly unknown mathematical structure.}
Below I give a logical path towards string theory starting from the Standard
Model. Pluses indicated successes, whereas minuses indicate problems which
force us to keep going down the list.

\begin{itemize}
\item Standard Model:
\begin{description}
\item[$+$] Experimentally very successful (Higgs boson?).
\item[$-$] Many unexplained features: $N_g=3$, fermion masses, quark mixings,
...
\end{description}
\item Unified theories:
\begin{description}
\item[$+$] Answer some of the Standard Model puzzles (charge quantization,
fermion mass relations, neutrino masses).
\item[$-$] Gauge hierarchy problem: why is $M_W\ll M_U$?
\end{description}
\item Supersymmetry:
\begin{description}
\item[$+$] Solves gauge hierarchy problem.
\item[$-$] Supersymmetry breaking: why should sparticles be $\lsim\TeV$
so that the gauge hierarchy problem is not reintroduced?
\end{description}
\item Supergravity:
\begin{description}
\item[$+$] Supersymmetry breaking becomes calculable in terms of few inputs.
\item[$-$] What determines the inputs to the supergravity model? What about
quantum gravity?
\end{description}
\item Superstrings:
\begin{description}
\item[$+$] Finite theory of quantum gravity, {\em everything} predicted in a
given vacuum (``model").
\item[$?$] Many possible vacua. What selects the ``correct" vacuum?
\end{description}
\end{itemize}

\subsection{What are strings?}
As a way of introduction, let us list several string characteristics:
\begin{itemize}
\item One-dimensional extended objects: $\sim10^{-33}$ cm in length.
\item Particles are identified with various string modes:
\begin{itemize}
\item massless modes which should contain the Standard Model particles, and
\item infinite tower of massive modes ($\sim M_{Pl}$ and higher).
\end{itemize}
\item Rather stringent consistency conditions (conformal invariance, modular
invariance) restrict the type of allowed string theories: bosonic, heterotic,
type II.
\item Heterotic string: closed string theory with two different choices
for the left- and right-moving string modes (supersymmetric and
non-supersymmetric). Most promising phenomenologically.
\item Number of degrees of freedom on the string is constrained. In the
simplest case consistency requires:
\begin{itemize}
\item non-supersymmetric string lives in 26 dimensions, and
\item superstring lives in 10 dimensions.
\end{itemize}
\item For the heterotic string in 10 dimensions, only two gauge groups are
allowed: $E_8\times E_8$ and $SO(32)$.
\item Must ``compactify" extra six dimensions (Calabi-Yau manifolds,
orbifolds), or can construct theories directly in four dimensions.
\item There are {\em many} ($\infty?$) four-dimensional models (compactified or
not). All these are allowed vacua of the theory.
\item Precise model-building rules give gauge group, spectrum, and interactions
for any given vacuum or ``model". This subject is called  ``string
phenomenology".
\end{itemize}

\section{String basics}
\begin{itemize}
\item The one-dimensional string sweeps out a two-dimensional world-sheet
embedded in $D$-dimensional spacetime.\footnote{For a textbook introduction
to string theory see Ref.~\cite{GSW}.}
\item The two-dimensional action describes the dynamical evolution in terms of
bosonic ($X^\mu$) and fermionic ($\psi^\mu$) fields on the world-sheet.
\item The classical solutions to the string equations of motion can be expanded
in terms of ``left-moving" and ``right-moving" modes. The two sectors are
basically decoupled (except that they must contribute equally to the mass of
the string) and can be chosen to be different theories (\ie, with or
without world-sheet supersymmetry):
\begin{itemize}
\item Bosonic string: non-supersymmetric$\otimes$non-supersymmetric,
\item Type II string: supersymmetric$\otimes$supersymmetric,
\item Heterotic string: supersymmetric$\otimes$non-supersymmetric.
\end{itemize}
\item The two-dimensional free-field action is conformal invariant at classical
level. However, conformal anomalies appear after quantization, with each
world-sheet field contributing a specific amount to the anomaly:
\begin{itemize}
\item each boson: $c=1$
\item each fermion: $c={1\over2}$
\item Fadeev-Popov ghosts: $c=-26$
\item Fadeev-Popov superghosts: $c=11$
\end{itemize}
The Fadeev-Popov ghosts appear in the quantization of the non-supersymmetric
string, whereas the superghosts appear additionally in quantizing the
supersymmetric string. In a consistent string theory the total contribution
to the conformal anomaly ($c_{\rm tot}$) must vanish.
\item Bosonic string:\\
 $c_{\rm tot}=1\cdot D^b_c+(-26)=0\Rightarrow D^b_c=26$ $X^\mu$ fields
required, \ie, the bosonic string lives in 26 dimensions.
\item Fermionic string: $c_{\rm tot}=1\cdot D^f_c+{1\over2}\cdot
D^f_c+(-26)+11=0
\Rightarrow D^f_c=10$ $(X^\mu,\psi^\mu)$ pairs required, \ie, the heterotic
and type II strings live in 10 dimensions.
\item If some of the degrees of freedom on the world-sheet are interpreted as
``internal" (not spacetime), the actual required dimension is lower than the
``critical" dimension calculated above, \ie, $D<D_c$.
\item Figure~\ref{vertex} shows how external string states are conformally
mapped onto the world-sheet and are represented by vertex operators. These
operators encode all of the quatum numbers of the string state.
\item String perturbation theory is an expansion in the topology of the
two-dimensional world-sheet (see Fig.~\ref{topology}).
\item Topologically distinct surfaces have different number of handles
(``genus"). Thus, at each order in perturbation theory there is only one
string diagram. This in contrast with the large number of diagrams present at
high orders in regular quantum field theory.\footnote{This stringy property has
been exploited to compute complicated QCD processes in the Standard Model in a
much simplified way, by viewing QCD as a low energy limit of string theory
\cite{BK}.}
\item String scattering amplitudes are defined as a path integral over the
two-dimensional quantum field theory on the world-sheet, with insertions of
suitable vertex operators representing the particles being scattered. This
recipe takes into account automatically the infinite number of massive modes
which could be exchanged in the scattering process.
\item In modern language this corresponds to calculating correlation functions
of vertex operators in the two-dimensional conformal field theory.
\end{itemize}

\section{Free-fermionic formulation}
\begin{itemize}
\item The idea is to formulate string theory directly in four dimensions. This
requires additional degrees of freedom on the world-sheet to cancel the
conformal anomaly. In the free-fermionic formulation \cite{FFF} these are
chosen to be free world-sheet fermions (with $c={1\over2}$ each). For the
heterotic string:
\begin{itemize}
\item left-movers (supersymmetric): $1\cdot D+{1\over2}\cdot D+{1\over2}\cdot
n_L+(-26)+11=0\Rightarrow n_L=18$ left-moving real fermions.
\item right-movers (non-supersymmetric): $1\cdot D+{1\over2}\cdot
n_L+(-26)=0\Rightarrow n_R=44$ right-moving real fermions.
\end{itemize}
\item The two-dimensional world-sheet fields are:
\begin{itemize}
\item left-movers: $X^\mu$, $\psi^\mu$, $(\chi^i,y^i,w^i)^{i=1\to6}$
\item right-movers: $\bar X^\mu$, $\bar\phi^{1\to44}$
\end{itemize}
\item States in the Hilbert space are constructed by acting on the vacuum
with creation operators associated with the ordinary ($X^\mu,\psi^\mu,\bar
X^\mu$) and free-fermionic world-sheet fields, \eg, $\psi^\mu_{1\over2}\bar
X^\nu_1|0\rangle$.
\item The frequencies of the creation operators determine the mass of the
string state, and their nature depends on the boundary conditions of the free
fermions as they are parallel transported around loops on the (one-loop)
world-sheet (torus). In the simplest case these boundary conditions can be
periodic or antiperiodic. Periodic boundary conditions imply integer
frequencies (``Ramond sector"); whereas antiperiodic boundary  conditions imply
half-integer frequencies (``Neveu-Schwarz sector").
\item The partition function of the world-sheet fermions depends on these
boundary conditions (these are called ``spin structures"). Constraints on the
spin structures follow by demanding world-sheet {\em modular invariance}, \ie,
physics should be independent of the two-dimensional surfaces being cut and
reconnected. Spin structures are specified for each world-sheet fermion and are
all collected in ``vectors" with 2+18 left-moving and 44 right-moving entries.
In these vectors $1\,(0)$ entries represent periodic (antiperiodic) boundary
conditions.
\item Further constraints imply that certain states must be dropped from the
spectrum. These states are eliminated by a set of generalized ``GSO
projections".
\item A consistent vacuum or ``model" is specified by:
\begin{itemize}
\item A basis for the spin-structure vectors $\{b_1,b_2,\ldots,b_n\}$.
\item An $n\times n$ matrix of GSO projections $C\left(\begin{array}{c}b_i\\
b_j\end{array}\right)$.
\item One must also verify that several consistency conditions are satisfied
for both allowed basis vectors and allowed GSO projection matrices.
\item The basis vectors span a space of ``sectors" of the Hilbert space,
which contain the allowed physical states. For a given state, the GSO
projection may or may not project it out.
\item For a state in a given sector $\alpha$ (a linear combination of the basis
vectors) its mass is given
by $M^2=-1/2+(1/8)\alpha^2_L+N_L=-1+(1/8)\alpha^2_R+N_R$, where $N_L(N_R)$
is the sum of the frequencies of the left-(right-)moving oscillators which
create the state and $\alpha^2_{L(R)}$ is the length-squared of the
left-(right-) moving part of $\alpha$.
\end{itemize}
\item These model-building rules lead to numerous possible models:
\begin{itemize}
\item the $b_i$ have 22+44 entries,
\item the $n\times n$ GSO-projection matrix has $n(n-1)/2$ independent
elements,
\item a typical model consists of $n=8$ vectors, \ie, $2^{(8\cdot7)/2}=268$
million choices.
\item There is a large amount of redundancy in the models so constructed.
Morever, many possibilities are ruled out on phenomenological grounds,
\eg, no spacetime supersymmetry, more or less than three generations of
quarks and leptons, no gauge group than can be broken down to the Standard
Model, etc.
\end{itemize}
\end{itemize}

\section{General results in free-fermionic models}
The free-fermionic formulation described above allows one to construct in
a systematic way large numbers of string models. The phenomenological
properties of these models can vary a lot from model to model, although there
are many models with nearly identical properties. We now list a few properties
which are generic in large classes of models of this kind.

\subsection{Gravity is always present}
\begin{itemize}
\item The simplest basis contains only one vector:
$\{b_1=${\bf1}$\}$,
 \ie, all fermions are periodic.
\item There are two sectors: $b_1$ which only contains massive states, and
$2b_1=0$ which is the Neveu-Schwarz sector.
\item Massless spectrum ($M^2=0$):
\begin{center}
\begin{tabular}{ll}
$\psi^\mu_{1\over2}\bar X^\nu_1|0\rangle_0$&
$\left\{\begin{array}{l}
{\rm graviton}\\ {\rm dilaton}\\ {\rm antisymmetric\
tensor}\end{array}\right.$\\
$\psi^\mu_{1\over2}\bar\phi^a_{1\over2}\bar\phi^b_{1\over2}|0\rangle_0$&
gauge bosons of $SO(44)$\\
$(\chi^i_{1\over2},y^i_{1\over2},w^i_{1\over2})\bar X^\mu_1|0\rangle_0$&
gauge bosons of $SU(2)^6$\\
$(\chi^i_{1\over2},y^i_{1\over2},w^i_{1\over2})\bar\phi^a_{1\over2}\bar\phi^b_{1\over2}|0\rangle_0$& scalars in adjoint of $SU(2)^6\times SO(44)$
\end{tabular}
\end{center}
These states are all allowed by the GSO projections. Therefore, the graviton
is {\em always} present in this type of models. Furthermore, gauge interactions
are also generically present.
\item There is also a tachyon: $\bar\phi^a_{1\over2}|0\rangle_0$ (with
$M^2<0$), since this is not a supersymmetric model (there are no massless
fermions).
\item The vacuum state $|0\rangle_0$ is the non-degenerate, spin-0,
Neveu-Schwarz vacuum.
\end{itemize}

\subsection{Spacetime supersymmetry}
\begin{itemize}
\item The next-to-simplest model has basis $\{b_1,S\}$, with
$$\begin{array}{cccccccccccccccccccc}
S&=&\psi^\mu&\chi^1&y^1&w^1&\chi^2&y^2&w^2&\cdots&\chi^6&y^6&w^6\\
&&1&1&0&0&1&0&0&\cdots&1&0&0\\
\end{array}$$
and $0$ for all the right-movers.
\item There are four sectors: $b_1$, $b_1+S$, $0$, $S$.
\item $S$ is the supersymmetry ``generator": the states in $b_1+S$ are the
superpartners of those in $b_1$ (all are massive), whereas the states
in $S$ are the superpartners of the Neveu-Schwarz states in $0$.
\item Massless spectrum:
\begin{center}
\begin{tabular}{ll}
\begin{tabular}{lcc}
&spin&\#\\
$\psi^\mu_{1\over2}\bar X^\nu_1|0\rangle_0$&2&1\\
$\bar X^\mu_1|0\rangle_S$&\begin{tabular}{c}$1\over2$\\ $3\over2$\end{tabular}
&\begin{tabular}{c}4\\ 4\end{tabular}\\
$\chi^i_{1\over2}\bar X^\mu_1|0\rangle_0$&1&6\\
\end{tabular}
&\begin{tabular}{l}
N=4 supermultiplet\\
(4 gravitinos)
\end{tabular}
\end{tabular}\\
\begin{tabular}{ll}
\begin{tabular}{lcc}
&helicity&\#\\
$\psi^\mu_{1\over2}\bar\phi^a_{1\over2}\bar\phi^b_{1\over2}|0\rangle_0$
&\begin{tabular}{c}$+1$\\ $-1$\end{tabular}
&\begin{tabular}{c}1\\ 1\end{tabular}\\
$\bar\phi^a_{1\over2}\bar\phi^b_{1\over2}|0\rangle_S$
&\begin{tabular}{c}$+{1\over2}$\\ $-{1\over2}$\end{tabular}
&\begin{tabular}{c}4\\ 4\end{tabular}\\
$\chi^i_{1\over2}\bar\phi^a_{1\over2}\bar\phi^b_{1\over2}|0\rangle_0$&0&6\\
\end{tabular}
&\begin{tabular}{l}
gauge supermultiplet\\
in adjoint of $SO(44)$
\end{tabular}
\end{tabular}
\end{center}
The model contains the complete $N=4$ supermultiplet with four gravitinos,
thus the model has $N=4$ supersymmetry. Also, the gauge particles form a
complete $N=4$ supermultiplet, as they should. (The multiplicity of states
in these multiplets is indicated in the $\#$ column.)
\item The $|0\rangle_S$ vacuum state is the degenerate, spin-1/2, Ramond
vacuum. This state is built from the eight periodic fermions in $S$ (\ie,
$\psi^\mu, \chi^{1\to6}$) which transform as the irreducible spinor
representation of $SO(8)$ (\ie, 4 Weyl fermions).
\item The vector $S$ brings in new GSO projections which {\em eliminate} the
tachyon from the spectrum, as it should be in a supersymmetric model.
\item With addition of further vectors to the basis, it is possible to reduce
the number of spacetime supersymmetries from 4 to 2 and then to 1. This is a
more complicated, although straightforward exercise.
\end{itemize}

\subsection{Gauge groups}
\begin{itemize}
\item The simply-laced Lie groups $SO(2n),SU(n),E_6,E_7,E_8$ are readily
obtainable. Others can also be obtained but in a less straightforward manner.
Usually one has to collect gauge boson states from several sectors to deduce
the gauge group.
\item A typical example is $SU(5)\times SO(10)\times SU(4)\times U(1)^n$.
\item In general, in first approximation all gauge couplings are {\em unified}
at the string scale $M_{\rm string}\approx5\times g\times10^{17}\GeV$, where
$g$ is the unified coupling \cite{Kaplunovsky}. String threshold corrections
can split the various gauge couplings by a model-dependent amount
\cite{thresholds}.
\item A very special property of string theory in general is that the gauge
coupling is determined dynamically as the expectation value of the dilaton
field $S$: $g^2\propto 1/\vev{S}$.
\item Models built in any formulation consist of sets of matter representations
which are automatically anomaly free. In practice this is a good check of the
derivation of complicated models.
\end{itemize}

\subsection{Superpotential}
\begin{itemize}
\item In the free-fermionic formulation the superpotential is calculable to any
order in the string fields by using the techniques of conformal field theory
\cite{KLN}. Generally one obtains
\begin{itemize}
\item Cubic terms: $\lambda\phi_1\phi_2\phi_3$, with $\lambda=c_3g$ and
$c_3=\{{1\over2},{1\over\sqrt{2}},1\}$.
\item Non-renormalizable terms: $\lambda\phi_1\phi_2\ldots\phi_N{1\over
M^{N-3}}$, with $\lambda=c_Ng$ and $c_N\sim1$ a calculable coefficient, and
$M\approx10^{18}\GeV$.
\end{itemize}
\item Non-trivial calculational techniques are required for the
non-renormalizable terms because of the coupling between left- and right-moving
degrees of freedom on the world-sheet. This property is related to the
asymmetric orbifold character of this class of models.
\item Non-renormalizable terms provide a natural hierarchical fermion mass
scenario \cite{decisive}:
\[
\begin{array}{ll}
\lambda Q_3 t^c H,& \lambda_t\sim g\, ;\\
\lambda Q_2 c^c H {\vev{\phi}\over M},&\lambda_c\sim g{\vev{\phi}\over M}\, ;\\
\lambda Q_1 u^c H {\vev{\phi}^2\over M^2},&\lambda_u\sim g{\vev{\phi}^2\over
M^2}\, .\\
\end{array}
\]
If $\vev{\phi}/M<1$, as is motivated by the cancellation of an anomalous
$U_A(1)$ symmetry always present in these models, then a hierarchy of Yukawa
couplings can be obtained.
\end{itemize}
\newpage

\section{Generic conformal field theory properties}
\begin{itemize}
\item All spacetime symmetries in string theory have their origin in
world-sheet symmetries.
\item Kac-Moody algebras: \cite{GO}
\begin{itemize}
\item The affine Kac-Moody algebra $\widehat G$ underlies the spacetime gauge
symmetry $G$. This algebra is represented by currents made of world-sheet
fermions, and can be realized at different positive integer {\em levels} $k$.
All states in the theory fall into representations of this algebra.
\item For a fixed level $k$, only certain representations to the gauge group
($G$) are unitary and thus allowed. These representations must satisfy
\[
\sum_{i=1}^{\rm rank\, G}n_i m_i\le k\, ,
\]
where the $n_i$ are the Dynkin labels of the highest weight representation, and
the $m_i$ are sets of numbers that depend on the gauge group,
\begin{center}
\begin{tabular}{|l|l|} \hline
&\hfil $m_i$\\ \hline
$SO(2n)$&$(1,2,2,\ldots,2,1,1)$\\
$SU(n)$&$(1,1,1,\ldots,1)$\\
$E_6$&$(1,2,3,2,1,2)$\\
$E_7$&$(2,3,4,3,2,1,2)$\\
$E_8$&$(2,3,4,5,6,4,2,3)$\\ \hline
\end{tabular}
\end{center}

\item Kac-Moody algebras (\ie, the degrees of freedom that they represent)
contribute to the central charge $c$ of the theory
\[
c={k\,{\rm dim}\, G\over k+\tilde h}\ ,
\]
where $\tilde h={1\over2}C_A$ is the dual Coxeter number, and $C_A$ the
quadratic Casimir of the adjoint representation.
\item For the non-supersymmetric right-movers, the conformal anomaly
cancellation equation is $4+c_{\rm matter}+(-26)=0
\Rightarrow c_{\rm matter}=22$. For a product group $G=\prod_i G_i$, the
constraint is stronger: $\sum_i c_i=22$.
\item Since the right-movers are responsible for representing the gauge group
of the string model, the corresponding Kac-Moody algebra should have a central
charge with $c\le22$. This entails an upper bound on the allowed level
\[ k\le \left[22\tilde h/({\rm dim}\,G-22)\right].
\]
For various groups of interest we get
\begin{center}
\begin{tabular}{ll}
$SU(5)$&$k\le55$\\
$SO(10)$&$k\le7$\\
$E_6$&$k\le4$\\
$E_7$&$k\le3$\\
$E_8$&$k\le2$
\end{tabular}
\end{center}
\end{itemize}
\item An allowed representation ($r$) at level $k$ is {\em massless} if
$h_r\le1$, where
\[
h_r={C_r\over 2k+C_A}
\]
is its {\em conformal dimension} ($C_r$ is the quadratic Casimir of the
representation\footnote{The group theoretical constants mentioned here ($n_i$,
$C_A$, $C_r$, etc.) have been tabulated \cite{Slansky}.}).
\item Combining the constraints from unitarity and masslessness, significant
restrictions follow on the allowed gauge group representations \cite{ELN}.
\end{itemize}

\begin{itemize}
\item Unitary massless representations at level 1
\begin{itemize}
\item $SO(2n)$: singlet, vector, and spinor. Spinor massless for\\ $n\le8$
only.
\item $SU(n)$: totally antisymmetric representations (see table).
\begin{center}
\begin{tabular}{|l|l|}\hline
$n$&Representation\\ \hline
2&\r{1},\r{2}\\
3&\r{1},\r{3},\rb{3}\\
4&\r{1},\r{4},\rb{4},\r{6}\\
5&\r{1},\r{5},\rb{5},\r{10},\rb{10}\\
6&\r{1},\r{6},\rb{6},\r{15},\rb{15},\r{20}\\
7&\r{1},\r{7},\rb{7},\r{21},\r{21},\r{35},\rb{35}\\
8&\r{1},\r{8},\rb{8},\r{28},\rb{28},\r{56},\rb{56},\r{70}\\
9&\r{1},\r{9},\rb{9},\r{36},\rb{36},\r{84},\rb{84}\\
10--23&\r{1},\r{n},\rb{n},{\bf n(n--1)/2},\rb{n(n-1)/2}\\ \hline
\end{tabular}
\end{center}
\vspace{0.5cm}
\item $E_6$: \r{1}, \r{27}, \rb{27}
\item $E_7$: \r{1}, \r{56}
\item $E_8$: \r{1}
\end{itemize}
\item Note that there are no adjoint representations allowed at level 1.
\end{itemize}

\begin{itemize}
\item What levels would be required for traditional GUT model building?
\begin{itemize}
\item $SU(5)$: for doublet-triplet splitting through the missing partner
mechanism one requires \r{50},\rb{50},\r{75} (the \r{75} breaks the gauge
symmetry). These are unitary at level $k\ge2$, but massless for $k\ge4$. The
unitary and massless representations at $k=4$ are:
\r{1},\r{5},\rb{5},\r{10},\rb{10},\r{15},\r{24},\r{40},\rb{40},\r{45},\rb{45},
\r{50},\rb{50},\r{75}. Therefore, very likely lots of exotics in the models.
\item $SO(10)$: requires \r{10},\r{16},\r{45},\r{54},\r{126},\rb{126}
\cite{BB}. All these are unitary and massless for $k\ge5$, but so are the
\r{144},\rb{144},\r{210}.
\end{itemize}
\item String model-building using level-one Kac-Moody algebras includes
almost every string model ever built. Model-building using higher-level
Kac-Moody algebras has been very limited because of the technical difficulties
involved \cite{Lewellen}. The proliferation of exotic representations is also a
potential problem.
\item At level-one, the $SU(5)\times U(1)$ gauge group becomes singled out
because the gauge symmetry is broken by \r{10},\rb{10} representations
\cite{revitalized}, which are allowed and occur in all known models of this
kind. There exist also string constructions in the free-fermionic formulation
where the Pati-Salam gauge group $SU(4)\times SU(2)\times SU(2)$ is obtained
\cite{ALR}, or even the Standard Model gauge group itself \cite{FNY,Alon}.
\end{itemize}

\section{SU(5)xU(1) GUT model-building}
{}From the previous section we conclude that $SU(5)\times U(1)$ is an
interesting
candidate for a string-derived gauge group. This gauge group is also quite
attractive from the traditional GUT (non-string) model-building perspective,
as we know recollect.
\begin{itemize}
\item The Higgs and matter fields are in the following representations:
\begin{itemize}
\item \r{10}: $H=\{Q_H,d^c_H,\nu^c_H\}$, \rb{10}: $\bar H=\{Q_{\bar
H},d^c_{\bar H},\nu^c_{\bar H}\}$,
\item \r{5}: $h=\{H_2,H_3\}$, \rb{5}: $\bar h=\{\bar H_2,\bar H_3\}$.
\item \r{10}: $F_i=\{Q,d^c,\nu^c\}_i$, \rb{5}: $\bar f_i=\{L,u^c\}_i$,
 \r{1}: $l_i=e^c_i$.
\end{itemize}
\item  The GUT superpotential is assumed to be:
\[
W_G=H\cdot H\cdot h+\bar H\cdot\bar H\cdot\bar h+F\cdot\bar H\cdot\phi
+\mu h\bar h,
\]
where the vacuum expectation values of the neutral components of the $H$ and
$\bar H$ fields ($\vev{\nu^c_H}=\vev{\nu^c_{\bar H}}=M_U$) break $SU(5)\times
U(1)$ down to $SU(3)\times SU(2)\times U(1)$.
\item Doublet-triplet splitting: the Higgs pentaplets ($h,\bar h$) have two
components with very different purposes
\[
h=\left(
\begin{array}{c}
H_2\\
H_3
\end{array}
\right)
\begin{array}{l}
{\rm electroweak\ symmetry\ breaking}\\
{\rm proton\ decay}
\end{array}
\]
The mass splitting of the doublets and triplets is accomplished by the
following superpotential interactions
\begin{eqnarray}
&H\cdot H\cdot h\to d^c_H\vev{\nu^c_H}H_3\nonumber\\
&\bar H\cdot\bar H\cdot\bar h\to \bar d^c_H\vev{\bar\nu^c_H}\bar H_3\nonumber
\end{eqnarray}
whereby the triplets get heavy, whereas the doublets remain light. This
phenomenon is a manifestation of the ``missing partner mechanism" \cite{MPM}.
(A similar mechanism in $SU(5)$ requires the introduction of large
representations for this sole purpose.)
\item The Yukawa part of the superpotential is given by
\[
\lambda_d F\cdot F\cdot h+\lambda_u F\cdot\bar f\cdot \bar h+\lambda_e
\bar f\cdot l^c\cdot h
\]
and generates the fermion masses. Note that unlike $SU(5)$, there is no
$m_b=m_\tau$ relation in $SU(5)\times U(1)$.
\item Neutrino masses follow from a generalized see-saw mechanism:
\[
\left.
\begin{array}{l}
F\cdot \bar f\cdot h\to m_u\nu\nu^c\\
F\cdot \bar H\cdot\phi\to \vev{\nu^c_{\bar H}}\nu^c\,\phi
\end{array}
\right\}
M_\nu=
\begin{array}{c}
\nu\\ \nu^c\\ \phi
\end{array}
\stackrel{\begin{array}{ccc} \nu\quad&\nu^c&\quad\phi\end{array}}
{\left(
\begin{array}{ccc}
0&m_u&0\\ m_u&0&M_U\\ 0&M_U&-
\end{array}
\right)}
\]
This mechanism gives $m_{\nu_{e,\mu,\tau}}\sim m^2_{u,c,t}/M_U$, which
have been shown to be consistent with the MSW mechanism, $\nu_\tau$ dark
matter, and ($\nu^c$) baryogenesis \cite{chorus+ENO+ELNO}.
\item Proton decay through dimension-six operators is mediated by heavy gauge
bosons, and is highly suppressed since $M_U\sim10^{18}\GeV$.
\item Proton decay mediated by dimension-five operators (see Fig.~\ref{d5pd})
\[
\lambda_d F\cdot F\cdot h\supset QQH_3\qquad
\lambda_ u F\cdot\bar f\cdot\bar h\supset QL\bar H_3
\]
is very suppressed since no $H_3,\bar H_3$ mixing exists, even though $H_3,\bar
H_3$ are heavy via the doublet-triplet splitting mechanism \cite{faspects}.
\end{itemize}
\newpage

\section{SU(5)xU(1) string model-building}
We now turn to the actual string models containing the gauge group $SU(5)\times
U(1)$ which have been built within the free-fermionic formulation.
\begin{itemize}
\item There are two variants of the model:
\begin{itemize}
\item the ``revamped" model, Antoniadis-Ellis-Hagelin-Nanopoulos (1989)
\cite{revamped}, and
\item the ``search" model, Lopez-Nanopoulos-Yuan (1992) \cite{search}.
\end{itemize}
These models have the following gauge group
\[
\underbrace{SU(5)\times U(1)}_{observable}\times
\underbrace{U(1)^n}_{mixed}\times\underbrace{SO(10)\times SU(4)}_{hidden}
\]
with $n=4\,(5)$ for the ``revamped" (``search") model.
\item The observable sector spectrum is schematically given by
\begin{flushleft}
\begin{tabular}{lll}
{\bf ``Revamped"}\\
\r{10}&4x&$\overbrace{10_f,10_f,10_f}^{3\ gens.},10_H$\\
\rb{10}&1x&$\overline{10}_H$\\
\rb{5}&4x&$\underbrace{\bar 5_f,\bar 5_f,\bar 5_f}_{3\ gens.},
\bar 5_{\rm EV}$\\
\r{5}&1x&$5_{\rm EV}$
\end{tabular}

\begin{tabular}{lll}
{\bf ``Search"}\\
\r{10}&5x&$\overbrace{10_f,10_f,10_f}^{3\ gens.},10_H,10_{\rm EV}$\\
\rb{10}&2x&$\overline{10}_H,\overline{10}_{\rm EV}$\\
\rb{5}&3x&$\underbrace{\bar 5_f,\bar 5_f,\bar 5_f}_{3\ gens.}$\\
\end{tabular}
\end{flushleft}
(Both models also contain \r{1} representations.)
\item Representations labelled ``EV" contain new, vector-like, heavy
particles, beyond those in the minimal supersymmetric standard model (MSSM).
\item Motivation for the ``search" model:
\begin{itemize}
\item the ``revamped" model unifies at best at $M_U\sim10^{16}\GeV$
\item the ``search" model has new $Q,\bar Q$ and $D^c,\bar D^c$ representations
to push $M_U$ up to the string scale:
\[
{\rm if}\ M_U\sim10^{18}\GeV,\ {\rm then} \left\{
\begin{array}{l}
M_{Q,\bar Q}\sim10^{12}\GeV\\
M_{D^c,\bar D^c}\sim10^6\GeV
\end{array}
\right.
\]
\end{itemize}

\item In general, to ``work out" a model one needs to study:
\begin{itemize}
\item the Higgs doublet mass matrix,
\item the Higgs triplet mass matrix,
\item the $F$- and $D$-flatness constraints in the presence of the anomalous
$U_A(1)$,
\item the dimension-five proton decay operators which may be reintroduced
in the string models.
\end{itemize}
\item Things are not simple because models contain many singlet fields,
which can (and some must) get vacuum expectation values (vevs). Moreover, many
entries in the Higgs mass matrices depend on these unknown vevs. Nonetheless,
it is possible to have all pieces of the model ``work out" for some choices of
the vevs. The resulting model is a deformation of the original free-fermionic
model.
\item Let us now describe one complete model (the ``search" model) starting
from the inputs (basis vectors and matrix of GSO projections) and giving the
results for the massless spectrum, the superpotential, and the Higgs mass
matrices.
\end{itemize}

\begin{itemize}
\item Basis vectors: the first
entry corresponds to the complexified $\psi^\mu$ and the next 18 entries to the
six left-moving triplets $(\chi^\ell,y^\ell,\omega^\ell)$. The first 12
right-moving entries (to the right of the colon) correspond to the real
fermions $\bar y^\ell,\bar\omega^\ell$, and the last 16 entries correspond to
complex fermions. A 1 (0) stands for periodic (antiperiodic) boundary
conditions. We also use the symbols $1_8=11111111$, $0_8=00000000$,
$A=\h\h\h\h1100$.
{
\begin{eqnarray}
{\bf 1}&=(1\ 111\ 111\ 111\ 111\ 111\ 111\ :\ 111111\ 111111\ 11111\ 111
\ 1_8)\nonumber\\
S&=(1\ 100\ 100\ 100\ 100\ 100\ 100\ :\ 000000\ 000000\ 00000\ 000\ 0_8)
\nonumber\\
b_1&=(1\ 100\ 100\ 010\ 010\ 010\ 010\ :\ 001111\ 000000\ 11111\ 100\
0_8)\nonumber\\
b_2&=(1\ 010\ 010\ 100\ 100\ 001\ 001\ :\ 110000\ 000011\ 11111\ 010\
0_8)\nonumber\\
b_3&=(1\ 001\ 001\ 001\ 001\ 100\ 100\ :\ 000000\ 111100\ 11111\ 001\
0_8)\nonumber\\
b_4&=(1\ 100\ 100\ 010\ 001\ 001\ 010\ :\ 001001\ 000110\ 11111\ 100\
0_8)\nonumber\\
b_5&=(1\ 001\ 010\ 100\ 100\ 001\ 010\ :\ 010001\ 100010\ 11111\ 010\
0_8)\nonumber\\
\alpha&=(0\ 000\ 000\ 000\ 000\ 000\ 011\ :\ 000001\ 011001\ \h\h\h\h\h\
\h\h\h\ A)\nonumber
\end{eqnarray}
}
\vspace{1cm}
\item The matrix of GSO projections:
{
\[
k=\pmatrix{
2 &1 &2 &2 &2 &2 &2 &1\cr
1 &1 &1 &1 &1 &1 &1 &4\cr
2 &2 &2 &2 &2 &2 &2 &1\cr
2 &2 &2 &2 &2 &2 &2 &1\cr
2 &2 &2 &2 &2 &2 &2 &4\cr
2 &2 &2 &2 &2 &2 &2 &1\cr
2 &2 &2 &2 &2 &2 &2 &3\cr
2 &1 &1 &1 &1 &1 &2 &3\cr}
\]
}
\end{itemize}
\begin{itemize}
\item The massless spectrum (the charges under the five $U(1)$'s are as
indicated)
\begin{itemize}
\item Observable Sector:
\[
\begin{array}{lll}
F_0\,(-\h,0,0,-\h,0)& F_1\,(-\h,0,0,\h,0)\\
F_2\,(0,-\h,0,0,0)& \bar f_2\,(0,-\h,0,0,0)& l^c_2\,(0,-\h,0,0,0)\\
F_3\,(0,0,\h,0,-\h)& \bar f_3\,(0,0,\h,0,\h)& l^c_3\,(0,0,\h,0,\h)\\
F_4\,(-\h,0,0,0,0)& \bar F_4\,(\h,0,0,0,0)\\
\bar F_5\,(0,\h,0,0,0)& \bar f_5\,(0,-\h,0,0,0)&
                                                l^c_5\,(0,-\h,0,0,0)\\
h_1\,(1,0,0,0,0)& \bar h_1\,(-1,0,0,0,0)\\
h_2\,(0,1,0,0,0)& \bar h_2\,(0,-1,0,0,0)\\
h_3\,(0,0,1,0,0)& \bar h_3\,(0,0,-1,0,0)\\
h_{45}\,(-\h,-\h,0,0,0)& \bar h_{45}\,(\h,\h,0,0,0)
\end{array}
\]
\item Singlets:
\[
\begin{array}{ll}
\Phi_{12}\,(-1,1,0,0,0)& \bar\Phi_{12}\,(1,-1,0,0,0)\\
\Phi_{23}\,(0,-1,1,0,0)&\bar\Phi_{23}\,(0,1,-1,0,0)\\
\Phi_{31}\,(1,0,-1,0,0)&\bar\Phi_{31}\,(-1,0,1,0,0)\\
\phi_{45}\,(\h,\h,1,0,0)& \bar\phi_{45}\,(-\h,-\h,-1,0,0)\\
\phi^+\,(\h,-\h,0,0,1)&\bar\phi^+\,(-\h,\h,0,0,-1)\\
\phi^-\,(\h,-\h,0,0,-1)&\bar\phi^-\,(-\h,\h,0,0,1)\\
\phi_{3,4}\,(\h,-\h,0,0,0)&\bar\phi_{3,4}\,(-\h,\h,0,0,0)\\
\eta_{1,2}\,(0,0,0,1,0)&\bar\eta_{1,2}\,(0,0,0,-1,0)\\
\Phi_{0,1,3,5}\,(0,0,0,0,0)
\end{array}
\]

\item Hidden Sector:\\  $T$: \r{10} of $SO(10)$;
$D$: \r{6}, $\F$: \r{4}, $\Fb$: \rb{4} of $SU(4)$.\\
The $\F_i,\Fb_j$ fields carry $\pm1/2$ electric charges.
\[
\begin{array}{ll}
T_1\,(-\h,0,\h,0,0)\\
T_2\,(-\h,-\h,0,0,-\h)\\
T_3\,(-\h,0,\h,0,0)\\
\\
D_1\,(0,-\h,\h,\h,0)& D_2\,(0,-\h,\h,-\h,0)\\
D_3\,(-\h,0,\h,0,0)&D_4\,(-\h,-\h,0,0,\h)\\
D_5\,(0,-\h,\h,0,0)& D_6\,(0,\h,-\h,0,0)\\
&D_7\,(\h,0,-\h,0,0)\\
\\
\F_{1}^+\,(-\q,\q,-\q,0,-\h)& \F_{2}^-\,(\q,\q,-\q,0,\h)\\
\F_{3}^+\,(\q,-\q,-\q,0,\h)&\F_{4}^+\,(-\q,\tq,\q,0,0)\\
\F_{5}^+\,(-\q,\q,-\q,0,\h)&\F_{6}^+\,(-\q,\q,-\q,0,-\h)\\
\Fb_{1}^-\,(-\q,\q,\q,\h,-\h)& \Fb_{2}^-\,(-\q,\q,\q,-\h,-\h)\\
\Fb_{3}^-\,(\q,-\q,\q,0,-\h)&\Fb_{4}^-\,(-\q,\q,\q,0,-\h)\\
\Fb_{5}^+\,(-\q,-\q,\q,0,-\h)&\Fb_{6}^-\,(-\tq,\q,-\q,0,0)
\end{array}
\]
\end{itemize}
\item The cubic superpotential is given by:
{
\begin{eqnarray}
W_3=g\sqrt{2}&\biggl\{
&F_0F_1h_1+F_2F_2h_2+F_4F_4h_1+F_4\bar f_5\bar h_{45}+F_3\bar f_3\bar h_3
							\nonumber\\
            &+&\bar f_2 l^c_2 h_2+\bar f_5 l^c_5 h_2\nonumber\\
&+&\rt F_4\bar F_5\phi_3+\h F_4\bar F_4\Phi_0
                  +\bar F_4\bar F_4\bar h_1+\bar F_5\bar F_5\bar h_2\nonumber\\
&+&(h_1\bar h_2\Phi_{12}+h_2\bar h_3\Phi_{23}+h_3\bar h_1\Phi_{31}
                           +h_3\bar h_{45}\bar\phi_{45}+{\rm h.c.})\nonumber\\
&+&\h(\phi_{45}\bar\phi_{45}+\phi^+\bar\phi^++\phi^-\bar\phi^-
                        +\phi_i\bar\phi_i+h_{45}\bar h_{45})\Phi_3\nonumber\\
&+&(\eta_1\bar\eta_2+\bar\eta_1\eta_2)\Phi_0
		+(\phi_3\bar\phi_4+\bar\phi_3\phi_4)\Phi_5\nonumber\\
&+&(\Phi_{12}\Phi_{23}\Phi_{31}+\Phi_{12}\phi^+\phi^-
                +\Phi_{12}\phi_i\phi_i+{\rm h.c.})\nonumber\\
&+&T_1T_1\Phi_{31}+T_3T_3\Phi_{31}\nonumber\\
&+&D_6D_6\Phi_{23}+D_1D_2\bar\Phi_{23}+D_5D_5\bar\Phi_{23}+D_7D_7\bar\Phi_{31}
							\nonumber\\
&+&D_3D_3\Phi_{31}+\h D_5D_6\Phi_0+\rt D_5D_7\bar\phi_3\nonumber\\
&+&\F_4\Fb_6\bar\Phi_{12}+\h F_3\Fb_4\Phi_0+\h F_2\Fb_5\Phi_3
                        +\F_6\Fb_4\phi^+\nonumber\\
&+&\rt \F_5\Fb_4\phi_4+\F_1\Fb_2D_5+\F_2\Fb_4l^c_2\biggr\}\nonumber
\end{eqnarray}
}
\item The quartic superpotential is given by:
\begin{eqnarray}
W_4&=
&F_2\bar f_2\bar h_{45}\bar\phi_4+F_3\bar F_4 D_4D_6+F_3\bar F_5 D_4
D_7\nonumber\\
&+&l^c_3\Fb_3\Fb_6D_7+l^c_5\F_2\Fb_3\bar\phi_3
                       +\F_1\Fb_3(\phi^+\bar\phi_3+\bar\phi^-\phi_3)\nonumber\\
&+&\Fb_3\Fb_5D_7\bar\phi^-+\F_2\F_5D_3\phi^-+\F_2\F_6D_3\phi_4+\F_5\Fb_1D_2D_7
\nonumber\\
&+&\F_5\Fb_2D_1D_7+\F_3\Fb_3D_3D_6+\F_4\Fb_3D_4D_7+\F_5\Fb_4D_5D_7.
                                                        \nonumber
\end{eqnarray}
Calculable coefficients ($\lambda=c_4g/M$, $c\sim1$, $M=10^{18}\GeV$) have been
omitted from $W_4$ but can be calculated using the methods of Ref.~\cite{KLN}.
\end{itemize}

\begin{itemize}
\item The Higgs doublet mass matrix is given by:
\[
{\cal M}_2=\bordermatrix{
&\ov H_1&\ov H_2&\ov H_3&\ov H_{45}\cr
H_1&0&\Phi_{12}&\bar\Phi_{31}&0\cr
H_2&\bar\Phi_{12}&0&\Phi_{23}&0\cr
H_3&\Phi_{31}&\bar\Phi_{23}&0&\bar\phi_{45}\cr
H_{45}&0&0&\phi_{45}&\Phi_3\cr
L_2&0&0&0&V_2\bar\phi_4\cr
L_3&0&0&V_3&0\cr
L_5&0&0&0&V_4\cr}
\]
Note that ``Higgs" ($H$) and ``lepton" ($L$) doublets cannot be distinguished
in principle. This result in consistent with Ref.~\cite{Martin} where it is
shown that in $SU(5)\times U(1)$ $R$-parity is not automatically conserved.
However, phenomenological constraints require that the bottom portion of the
matrix decouples from the top portion \cite{search,LNZI} and therefore the
standard $R$-parity symmetry is present in the model.
\item The Higgs triplet mass matrix is given by:
\[
{\cal M}_3=\bordermatrix{
&\bar D_1&\bar D_2&\bar D_3&\bar D_{45}&d^c_0&d^c_1&d^c_2&d^c_3&d^c_4\cr
D_1&0&\Phi_{12}&\bar\Phi_{31}&0&V_1&V_0&0&0&V_4\cr
D_2&\bar\Phi_{12}&0&\Phi_{23}&0&0&0&V_2&0&0\cr
D_3&\Phi_{31}&\bar\Phi_{23}&0&\bar\phi_{45}&0&0&0&0&0\cr
D_{45}&0&0&\phi_{45}&\Phi_3&0&0&0&0&0\cr
\bar d^c_4&\ov V_4&0&0&0&w^{(4)}_0&w^{(4)}_1&0&0&\Phi_0\cr
\bar d^c_5&0&\ov V_5&0&0&w^{(5)}_0&w^{(5)}_1&0&0&\phi_3\cr}
\]
In this case it is automatic that three $d^c$ states remain light, but which
linear combinations these are is model dependent. In fact, mixings among the
``canonical" $d^c_{0,1,2,3,4}$ could be an important source of
Kobayashi-Maskawa mixing at low energies \cite{EVA}.
\end{itemize}
\newpage

\begin{itemize}
\item String scenario for a ``heavy" top quark:
\begin{itemize}
\item We identify $g\sqrt{2}F_4\bar f_5\bar h_{45}$ with the top-quark Yukawa
coupling, and get $\lambda_t(M_U)=g\sqrt{2}$.
\item At low energies one gets $m_t=\lambda_t(m_t) \sin\beta (174)\GeV$, where
$\lambda_t(m_t)$ is the top-quark Yukawa coupling at low energies and
$\tan\beta$ is the ratio of the Higgs vacuum expectation values.
\item In Fig.~\ref{yuks} we show the top-quark Yukawa coupling at the
unification scale versus $m_t$ (figure from Ref.\cite{t-paper})
for fixed values of $\tan\beta$. The horizontal lines indicate a possible
range of string predictions for the top-quark Yukawa coupling.
\item One can see that the experimentally preferred values of $m_t$
($\sim170\pm10 \GeV$) fit well with typical string predictions.
\end{itemize}
\end{itemize}

\section{Supersymmetry breaking}
Since the superpartners of the ordinary particles have not been observed,
supersymmetry must be a broken symmetry at low energies. The mechanism of
supersymmetry breaking remains unclear, although great strides in this
direction have been made in the last few years, especially with inspiration
from string theory. The most popular mechanism for dynamical supersymmetry
breaking can be outlined as follows:
\begin{itemize}
\item The hidden sector gauge group may become strongly interacting at some
intermediate scale depending on the hidden gauge group and the hidden matter
content.  Gaugino condensation will likely then occur, \ie,
$\vev{\lambda\lambda}\not=0$.
\item The scalar potential must have a minimum which breaks supersymmetry.
This appears to require a tuning of two different hidden sector gauge groups
with similar gauge and matter content \cite{susyx}. However, this may not be
necessary if the strong interactions are handled in a less than naive way
\cite{Ross}.
\item The breaking of supersymmetry may be due to the $F$-term of several
possible fields: the dilaton ($S$), the moduli ($T$), or the hidden matter
($H$) fields.
\item Since $\vev{S}\propto 1/g^2$, the scalar potential must have a minimum in
the $S$ direction for a finite value of $\vev{S}$. A typical problem is
$\vev{S}\to\infty$, \ie, $g\to0$.
\item The vacuum energy (value of the scalar potential at the minimum) is the
cosmological constant, and therefore should be ``small". There is no general
solution to this condition, although particular cases may just work out.
\item The magnitude of supersymmetry breaking in the observable sector should
not exceed $\lsim1\TeV$. This can be accomplished with suitably chosen hidden
sectors \cite{Casas}.
\end{itemize}

The above scenario for supersymmetry breaking is ``string-inspired" but not
necessarily consistent with string theory. For example, possible string
non-perturbative effects are ignored, and the solution to cosmological constant
problem is assumed not to impact the results.

I should also mention another scenario for supersymmetry breaking in string
theory, through the so-called Scherk-Schwarz mechanism. In this case
supersymmetry is broken perturbatively and its magnitude can come out to be
small enough if there is a modulus field which acquires a very large
expectation value. This is equivalent to an effective decompactification of
a compactified dimension and can be realistic if some conditions are
satisfied \cite{Ig}.

Recently a more model-independent approach to string-inspired supersymmetry
breaking has become popular \cite{KL,BIM}. In this approach supersymmetry
breaking is parametrized by an angle $\tan\theta=\vev{F_S}/\vev{F_T}$.
Generally one finds that the scalar masses are not universal
\[
 m^2_i=m^2_{3/2}(1+n_i\cos^2\theta),
\]
where $m_{3/2}$ is the gravitino mass, and $n_i$ is the modular weight of the
string state. Non-universality of the first and second-generation scalar masses
could easily violate stringent limits on flavor changing neutral currents in
the $K$-system \cite{EN}. If one demands universality of the scalar masses, two
scenarios arise:
\begin{description}
\item (i) $\cos\theta=0\Leftrightarrow\vev{F_S}\gg\vev{F_T}$: ``{\bf dilaton
scenario}" \cite{KL,BIM}
\[
m_i=m_{3/2},\quad M_a=\sqrt{3}m_{3/2},\quad A=-\sqrt{3}m_{3/2}
\]
\item (ii) All $n_i$ equal ($n_i=-1$). This occurs in $Z_2\times Z_2$ orbifold
models, free-fermionic models,\footnote{Detailed studies of this question in
realistic free-fermionic models are in progress \cite{LNY94}.}
 and in the large-$T$ limit of Calabi-Yau compactification. If
$\cos\theta=1\Leftrightarrow
\vev{F_T}\gg\vev{F_S}$, then $m_i=0$ and we call this the ``{\bf moduli
scenario}". More generally: $m_i=\sin\theta\, m_{3/2}$,
$M_a=\sqrt{3}\sin\theta\, m_{3/2}$, and $A=-\sqrt{3}\sin\theta\, m_{3/2}$. (If
$\sin\theta\to0$, one needs to worry about one-loop corrections to K\"ahler
potential and gauge kinetic function \cite{BIM}.)
\end{description}

\section{The bottom-up approach}

\begin{itemize}
\item The previous discussion has been mostly about true string phenomenology.
However, the subject of supersymmetry breaking already steps into the
``string-inspired" phenomenology area, although not completely.
\item We now depart from the rigorous string predictions and turn to {\em
string-inspired} phenomenology. In the present context, this consists of taking
the best known properties of string models and building a supergravity model
based on them. Eventually a single string model may be found where all the
desired properties may happen simultaneously. This model would be a true
candidate for a fundamental ``Theory of Everything".
\item Our string-inspired model consists of:
\begin{itemize}
\item An $SU(5)\times U(1)$ supergravity model whose gauge couplings unify at
the string scale $M_{\rm string}\sim10^{18}\GeV$. This requires one vector-like
quark doublet ($M_Q\sim10^{12}\GeV$) and one vector-like quark singlet
($M_D\sim10^6\GeV$), in addition to the particles in the Minimal Supersymmetric
Standard Model. These particles occur in the ``search" string $SU(5)\times
U(1)$ model of Ref.~\cite{search}.
\item We also assume that supersymmetry breaking is triggered by a set of
soft-supersymmetry-breaking terms which correspond to the ``moduli" and
``dilaton" universal soft supersymmetry breaking scenarios discussed above.
\end{itemize}
\end{itemize}

\subsection{Unification of gauge couplings}

As just mentioned, the unification of the gauge couplings at the string scale
require the existence of new intermediate-mass particles. Their masses
depend on the value of the strong coupling, as shown in the following table
\begin{center}
\begin{tabular}{|c|c|c|c|}\hline
$\alpha_3(M_Z)$&$M_{D}\,(\GeV)$&$M_{Q}\,(\GeV)$&$\alpha(M_U)$\\ \hline
$0.110$&$4.9\times10^4\GeV$&$2.2\times10^{12}\GeV$&$0.0565$\\
$0.118$&$4.5\times10^6\GeV$&$4.1\times10^{12}\GeV$&$0.0555$\\
$0.126$&$2.3\times10^8\GeV$&$7.3\times10^{12}\GeV$&$0.0547$\\ \hline
\end{tabular}
\end{center}

In figure~\ref{proc_erice1} we show the running of the gauge couplings (solid
lines) and their unification at the string scale $M_{\rm
string}\sim10^{18}\GeV$. For reference we also show the case of no
intermediate-scale particles (dotted lines) where the gauge couplings unify at
a lower scale.

\subsection{Soft supersymmetry breaking}
We now list all of the soft-supersymmetry-breaking parameters generally
allowed in supergravity models. The assumption of {\em universality} of the
soft parameters is also commonly made. This assumption has some basis in
the original supergravity models, but is not guaranteed, and in fact it is
explicitly violated in most string-inspired supersymmetry breaking models. We
also keep a running list of how many parameters are being introduced at each
stage.
\begin{itemize}
\item {\bf Gaugino masses} (parameters = 3)
\begin{itemize}
\item $M_3,M_2,M_1$ parametrize the masses of the superpartners of the
gauge bosons of $SU(3)_C,SU(2)_L,U(1)_Y$.
\item The universal soft-supersymmetry-breaking assumption entails:
$M_3=M_2=M_1=m_{1/2}$ at $M_U$. However, there is no phenomenological
reason for such requirement, as far as the gaugino masses are concerned.
\end{itemize}
\item {\bf Scalar masses} (parameters = $5\times3+2 = 17$)
\begin{itemize}
\item Need to provide masses for the squarks and sleptons\\ $(\tilde Q,\tilde
U^c,\tilde D^c,\tilde L,\tilde E^c)_i$, $i=1,2,3$, and for the
Higgs-boson doublets $H_1, H_2$.
\item Universality implies that {\em all} these masses are equal to $m_0$ at
$M_U$. Limits on flavor-changing-neutral-currents (FCNCs) at low energies
require that the squarks (and sleptons) of the first two generations be nearly
degenerate in mass \cite{EN}. The universality assumption together with the
renormalization group evolution of the scalar masses assures that the
experimental limits on FCNCs are easily satisfied. We note that there are
no comparable experimental constraints on the third generation sparticles or
on the Higgs-boson doublets; nonetheless universality is usually assumed for
these masses too.
\end{itemize}
\item {\bf Scalar couplings} (parameters = $3+1 = 4$)
\begin{itemize}
\item Each superpotential coupling is accompanied by a
soft-supersymmetry-breaking term proportional to it. These soft terms are
trilinear couplings among the scalar components of the superfields which appear
in the corresponding superpotential terms, as follows
\[
\begin{array}{ll}
\lambda_tQt^cH_2&\to \lambda_tA_t\tilde Q\tilde t^c H_2\\
\lambda_bQb^cH_1&\to \lambda_bA_b\tilde Q\tilde b^c H_1\\
\lambda_\tau L\tau^cH_1&\to \lambda_\tau A_\tau\tilde L\tilde\tau^c H_1\\
\mu H_1 H_2&\to \mu B H_1 H_2
\end{array}
\]
\item In this case the universality assumption entails: $A_t=A_b=A_\tau=A$ at
$M_U$.
\end{itemize}
\item All of the soft-supersymmetry-breaking parameters, and the
gauge and Yukawa couplings evolve down to low energies as prescribed by the
appropriate set of coupled renormalization group equations (RGEs).
\end{itemize}

\subsection{Parameter count at low energies}
The following is a list of all the parameters introduced in
supersymmetric models (excluding CP-violating phases). The ``MSSM" column
counts the number of parameters in the Minimal Supersymmetric Standard Model,
wherea the ``SUGRA" column refers to the case of supergravity models with
universal soft supersymmetry breaking.
\[
\begin{array}{lccc}
{\rm Parameter}&{\rm MSSM}&{\rm SUGRA}&\\
M_1,M_2,M_3&3&1&(m_{1/2})\\
(\tilde Q,\tilde U^c,\tilde D^c,\tilde L,\tilde E^c)_i&15&1&(m_0)\\
H_1,H_2&2&0&(m_0)\\
A_t,A_b,A_\tau&3&1&(A)\\
B&1&1&{\rm (determined\ by}\\
\mu&1&1&{\rm radiative\ EWx)}\\
\lambda_{b,t,\tau},\tan\beta&2&2&\\
&-&-&\\
{\rm Total:}&27&7&
\end{array}
\]
\begin{itemize}
\item The two minimization conditions of the electroweak scalar
potential impose two additional constraints which can be used to determine
$\mu,B$ and thus reduce the parameter count down to {\bf5} (versus 25 in the
MSSM).
\item In the two string-inspired scenarios we consider, $m_0$ and $A$ are known
functions of $m_{1/2}$, therefore the parameters are only {\bf3}.
\item Moreover, in a self-consistent supersymmetry breaking theory, even
$m_{1/2}$ (or the relevant scale) would be determined. With the knowledge of
$m_t$, only $\tan\beta$ would remain unknown.
\end{itemize}

\subsection{Determination of the theoretically allowed parameter space}
The theoretically allowed parameter space in the variables $(m_t,\tan\beta,
m_{1/2},m_0,A)$ can be determined by a self-consistent procedure of
running the RGEs for the various parameters between the weak scale and
unification scale and imposing the electroweak breaking constraint. This
procedure is non-trivial and is not new \cite{EWx,LN}. However, because of
the revival of supersymmetric grand unification, this procedure has been
re-examined in detail prior to the LEP era \cite{preLEP}, during the early
LEP years \cite{LEPera,aspects}, and also very recently \cite{recent,Layman}.
Here we just present a ``flow chart" of the various steps which are generally
followed. This is given in Fig.~\ref{chart}.

The various inputs in the calculation are:
\begin{itemize}
\item known quantities: $m_b,m_\tau,\alpha_3,M_Z$
\item the top-quark mass $m_t$
\item $\tan\beta$
\item the three universal soft-supersymmetry-breaking parameters $m_{1/2}$,
$m_0$, and $A$
\end{itemize}
Note that the parameters are input at different scales.
\newpage

\section{Radiative Electroweak Breaking}
\subsection{Tree-level minimization}
The tree-level Higgs potential, assuming that only the neutral
components get vevs, is given by
\begin{eqnarray}
V_0&=&(m^2_{H_1}+\mu^2)h_1^2+(m^2_{H_2}+\mu^2)h_2^2+2B\mu h_1h_2\nonumber\\
&&+\coeff{1}{8}(g_2^2+g'^2)(h^2_2-h^2_1)^2,\nonumber
\end{eqnarray}
where $h_i= {\rm Re}\,H^0_{1,2}$. One can then write down the minimization
conditions $\partial V_0/\partial h_i=0$ and obtain
\begin{eqnarray}
\mu^2&=&{m^2_{H_1}-m^2_{H_2}\tan^2\beta\over\tan^2\beta-1}
-\coeff{1}{2}M^2_Z,\nonumber\\
B\mu&=&-\coeff{1}{2}\sin2\beta(m^2_{H_1}+m^2_{H_2}+2\mu^2).\nonumber
\end{eqnarray}
The solutions to these equations are physically sensible only if they
reflect a minimum away from the origin
\[
{\cal S}=(m^2_{H_1}+\mu^2)(m^2_{H_2}+\mu^2)-B^2\mu^2<0
\]
of a potential bounded from below
\[
{\cal B}=m^2_{H_1}+m^2_{H_2}+2\mu^2+2B\mu>0.
\]
Taking the second derivative $\partial^2 V_0/\partial h_i\partial h_j$ one
can determine the physical tree-level Higgs masses.

\subsection{One-loop minimization}
\noindent The above procedure is however not completely satisfactory since
the tree-level scalar potential has minima which are not renormalization-scale
independent. Indeed, in Fig.~\ref{potential} we show the typical change
in shape of the scalar potential as the renormalization group scale is
lowered. This variation implies that the vacuum expectation values which
give the $Z$-boson mass vary {\em a lot} for scales $Q\lsim1\TeV$, as
shown schematically in Fig.~\ref{vevs}.

The problem is that  ${dV_0\over d\,\ln Q}\not=0$, that is, the tree-level
scalar potential does not satisfy the renormalization group equation, and
$Q$-dependence is present. The solution to this problem is to use instead the
one-loop effective potential
\[
V_1=V_0+\Delta V,\]
with
\[
\Delta V=\coeff{1}{64\pi^2}{\rm Str}\,{\cal M}^4\left(\ln{{\cal M}^2\over Q^2}
-\coeff{3}{2}\right),
\]
where ${\rm Str}{\cal M}^2=\sum_j(-1)^{2j}(2j+1){\rm Tr}{\cal M}^2_j$.
This potential satisfies ${dV_1\over d\,\ln Q}=0$ (to one-loop order) and
the $Q$-dependence can be minimized \cite{aspects}.

The vevs are now $Q$-independent (up to two-loop effects) in the range
of interest ($\lsim1\TeV$). However, one must perform the minimization of
the potential numerically (non-trivial) and {\em all} the spectrum enters into
$\Delta V$ (although $\tilde t,\tilde b$ are the dominant contributions).
This method also gives automatically the {\em one-loop corrected}
Higgs boson masses (taking second derivatives of $V_1$).

\subsection{Radiative Symmetry Breaking}
Let us now examine how the symmetry is actually broken in the simple
(and unrealistic) case of $\mu=0$ and just considering the tree-level
potential. In this case the ${\cal S}<0$ condition (minimum away from the
origin) reduces to
\[
{\cal S}\to m^2_{H_1}\cdot m^2_{H_2}<0,
\]
and we must arrange that one $m^2_{H}<0$ somehow.

Consider the relevant RGEs schematically (setting $\lambda_b=\lambda_\tau=0$)
\[
{d\widetilde m^2\over dt}={1\over(4\pi)^2}\left\{
-\sum_i c_i g^2_i M^2_i + c_t\lambda^2_t\left(\sum_i\widetilde
m^2_i\right)\right\}
\]
with coefficients
\[
\begin{array}{cccc}
&c_t&c_3&c_2\\
H_1&0&0&6\\
H_2&6&0&6\\
\widetilde Q&0&\coeff{32}{3}&6\\
\widetilde U^c&0&\coeff{32}{3}&0\\
\widetilde D^c&0&\coeff{32}{3}&0\\
\widetilde L&0&0&6\\
\widetilde E^c&0&0&0
\end{array}
\]
The runnings of the scalar masses are shown in Fig.~\ref{runnings} for a
particular choice of the parameters, although the qualitative result is
correct in most cases of interest. We observe:
\begin{itemize}
\item $m^2_{H_1}<0$, and $m^2_{H_2}<0$ for $Q<Q_0$
\item $m^2_{\tilde Q,\tilde U^c,\tilde D^c}>0$ because of the large $\alpha_3$
contribution in their RGEs, and the smaller $\lambda_t$ dependence.
\end{itemize}
Therefore, the electroweak symmetry is broken ``radiatively" and the
squark and slepton squared masses remain positive.

\section{The allowed parameter space}

\noindent We consider the two string-inpired soft-supersymmetry-breaking
scenarios discussed above:
\begin{itemize}
\item ``moduli" scenario $m_0=A=0$ \cite{LNZI}
\item ``dilaton" scenario $m_0={1\over\sqrt{3}}m_{1/2}$, $A=-m_{1/2}$
\cite{LNZII} (This scenario has also been considered in the context of the
MSSM in Ref.~\cite{BLM}.)
\end{itemize}
\noindent There are only three parameters:
\begin{itemize}
\item $m_{1/2}\leftrightarrow m_{\tilde g}\leftrightarrow m_{\chi^\pm_1}$
\item $\tan\beta$
\item $m^{\rm pole}_t\quad\left\{\begin{array}{ll} >131\GeV&{\rm D0}\
\cite{D0}\\ 160\pm13\GeV& {\rm EW\ fits\ for\ light\ Higgs}\ \cite{Altarelli}\\
174\pm17\GeV&{\rm CDF}\ \cite{CDF}\end{array}\right.$
\end{itemize}
\noindent Parameter space:
\begin{itemize}
\item We fix $m_t=150,170\GeV$, and vary $\tan\beta$ and $m_{\chi^\pm_1}$.
\item {\bf Note}: the ``pole" mass ($m_t^{\rm pole}$, as measured
experimentally) is 5\% higher than the ``running" mass ($m_t$) which we use
here. Therefore, our $m_t$ choices correspond to $m_t^{\rm
pole}\approx157,178\GeV$.
\item We keep only points which satisfy all LEPI bounds on sparticle and
Higgs-boson masses ($m_{\chi^\pm_1}>45\GeV$,
$\Gamma^{inv}_Z$, $m_h\gsim60\GeV$, $m_{\tilde l}\gsim45\GeV$) \cite{aspects}.
\item The lightest supersymmetric particle (LSP) is the lightest neutralino
$\chi$, with a calculated relic abundance $\Omega_\chi h^2_0<1$. Thus,
cosmological constraints are automatically satisfied.
\end{itemize}
\bigskip

The parameter spaces are shown in Fig.~\ref{moduliPS} and Fig.~\ref{dilatonPS}
for the moduli and dilaton scenarios respectively \cite{Easpects}. The allowed
points in parameter space are marked by various symbols. Excluded points are
blank. Further experimental constraints apply, as discussed below, and lead to
further excluded points (all points with symbols other than a period).

\section{Experimental constraints}
We now discuss further experimental constraints which restric the parameter
spaces in the moduli and dilaton scenarios.

\subsection{$b\to s\gamma$}
\begin{itemize}
\item There are three main contributions to this process (see Fig.~\ref{bsgd}):
(i) the $W-t$ loop, (ii) the $H^\pm-t$ loop, and the (iii) the $\chi^\pm-\tilde
t$ loop. The first two contributions are negative, whereas the last one could
have either sign. In fact, its sign is strongly correlated with the sign of
$\mu$.
\item In $SU(5)\times U(1)$ supergravity these contributions have been
calculated in Ref.~\cite{bsgamma+bsg-eps} and are shown in Fig.~\ref{bsgr}
for the moduli scenario. For $\mu>0$ one can observe the destructive
interference effects. The horizontal lines correspond to the latest CLEOII
limits  $B(b\to s\gamma)=(0.6-5.4)\times10^{-4}$ at 95\%CL \cite{Thorndike}.
\item One should be aware that one-loop QCD corrections change the tree-level
result by a large factor, thus two-loop QCD corrections are expected to be
large as well. The lack of a complete two-loop calculation is the largest
source of uncertainty in this calculation.
\item However, since supersymmetric contributions can be much larger or much
smaller than the Standard Model prediction (which depends only on $m_t$, see
$\mu<0$ in Fig.~\ref{bsgr}), a measurement of $B(b\to s\gamma)$ will (and
already has) constrain(ed) the parameter space in important ways. The excluded
points of parameter space are denoted by pluses ($+$) in
Figs.~\ref{moduliPS},\ref{dilatonPS}.
\end{itemize}

\subsection{$(g-2)_\mu$}
\begin{itemize}
\item The supersymmetric one-loop contributions to the anomalous magnetic
moment of the muon $(g-2)_\mu$ are shown in Fig.~\ref{g-2d}, and have been
calculated in Ref.~\cite{g-2}. The results in the moduli scenario are shown in
Fig.~\ref{g-2r}.
\item The present experimental value for $(g-2)_\mu$ is
$a^{exp}_\mu=1165923\,(8.5)\times10^{-9}$ \cite{oldg},
whereas the latest Standard Model prediction is
$a^{SM}_\mu=1165919.20\,(1.76)\times10^{-9}$ \cite{kinoII}. From these
two numbers we get an allowed interval (at 95\%CL) for any
beyond-the-standard-model contribution:
\[
-13.2\times10^{-9}<a^{susy}_\mu<20.8\times10^{-9}.
\]
\item The figure indicates that this limit is easily violated for not so small
values of $\tan\beta$. In fact, there is a significant enhancement in
$(g-2)_\mu$ for large $\tan\beta$ \cite{g-2}. Points presently excluded are
denoted by crosses ($\times$) in Figs.~\ref{moduliPS},\ref{dilatonPS}.
\item The new Brookhaven E821 experiment \cite{newg} expects to reach an
ultimate sensitivity of $0.4\times10^{-9}$, and is slated to start taking data
in January of 1996.
\item This new sensitivity is designed to test the electroweak contribution to
$(g-2)_\mu$, which is much smaller than the typical supersymmetric
contribution. Therefore we expect very important restrictions on the parameter
space of (or indirect evidence for) supersymmetric models.
\end{itemize}

\subsection{$\epsilon_1,\,\epsilon_b$}
\begin{itemize}
\item The one-loop electroweak corrections to the LEP observables can
be parametrized in terms of four quantities: $\epsilon_{1,2,3,b}$ \cite{ABC}.
Of these, only $\epsilon_1$ (related to the $\rho$-parameter) and $\epsilon_b$
(related to $Z\to b\bar b$) have been constraining at all over the running of
LEP. Both these parameters have a quadratic dependence on $m_t$; $\epsilon_1$
also has a logarithmic dependence on the Higgs-boson mass.
\item The main diagrams contributing to these parameters in supersymmetric
models are shown in Figs.~\ref{eps1} and \ref{epsb}. In both cases, the
effects of supersymmetry are most significant for a light chargino
$m_{\chi^\pm_1}\lsim60-70\GeV$, which shifts $\epsilon_1$ negatively and thus
compensates for a large top-quark mass. An example of this effect on
$\epsilon_1$ is shown in Fig.~\ref{eps1inset}. Otherwise supersymmetry
decouples completely and the Standard Model results with a light Higgs boson
are obtained.
\item In Fig.~\ref{Easpects5a} \cite{Easpects} we show the calculated values of
$\epsilon_1$ and $\epsilon_b$ in the $SU(5)\times U(1)$ model with moduli
scenario. The various experimental ellipses indicate a preference for lighter
top-quark masses. The size of the ellipses is expected to be reduced by a
factor of two with the 93+94 LEP data.
\end{itemize}

\subsection{Neutrino telescopes}

\begin{itemize}
\item Neutralinos ($\chi$) in the galactic halo are captured by the Sun and the
Earth and eventually annihilate $\chi\chi\to f\bar f\to\cdots\to$ into
high-energy $\nu$'s. These neutrinos can then travel to underground (or
underwater) detectors, such as Kamiokande, MACRO, Amanda, Nestor, Dumand.
\item The signal is that of upwardly-moving muon fluxes in the detector
which are above the expected atmospheric neutrino background. At present there
are only flux limits from Kamiokande, although limits from MACRO are
forthcoming.
\item The concentration of ${\rm Fe}^{56}$ nuclei on Earth enhances the capture
of neutralinos with mass close to 56 GeV. In Fig.~\ref{NT} \cite{NT} we
show the predicted flux for Earth capture and the present Kamiokande upper
limit. One can see the ${\rm Fe}^{56}$ enhancement and that the data already
impose some (although small) restrictions on parameter space. The excluded
points are denoted by diamonds ($\diamond$) in
Figs.~\ref{moduliPS},\ref{dilatonPS}.
\item It is expected that once MACRO is fully operational, an improvement
in flux sensitivity by a factor of 2--10 would be achieved.
\end{itemize}
\newpage

\section{Prospects for direct detection}
We now discuss the prospects for direct detection of the sparticles and
Higgs bosons in the string-inspired $SU(5)\times U(1)$ models which we
have discussed above. The parameter space which is explored is that which
is allowed by all of the experimental constraints introduced in the
previous section.

\subsection{Tevatron}
\begin{itemize}
\item The missing energy signature in $\tilde q,\tilde g$ production is
kinematically disfavored since one generally obtains $m_{\tilde q}\approx
m_{\tilde g}\gsim250\GeV$ in $SU(5)\times U(1)$ supergravity.
\item A lot more accessible is the {\em trilepton} channel \cite{trileptons}:
$p\bar p\to \chi^0_2\chi^\pm_1X\to 3l$. In these models one typically obtains
large leptonic branching fractions for the charginos $B(\chi^\pm_1\to
e+\mu)\approx2/3$, whereas $B(\chi^0_2\to ee,\mu\mu)$ can be small for light
charginos.
\item In Fig.~\ref{trilep} we show the rate for trilepton events versus the
chargino mass in the $SU(5)\times U(1)$ ``moduli scenario" (the results are
somewhat smaller in the ``dilaton scenario") \cite{LNWZ,Easpects}. The
solid line is the present CDF upper limit on the trilepton rate \cite{Kato}.
\item Experimental efficiencies for detection of trilepton events are
small ($<10\%$) because of large ``instrumental" backgrounds, \ie, when jets
``fake" leptons in the detector.
\item By the end of run IB (1993--95) it is expected that $\sim100\ipb$
of data would be collected by each detector. This should move the experimental
limit from the solid line in Fig.~\ref{trilep} down to the dashed line,
\ie, probing chargino masses as high as 100 GeV. Points in parameter space
reachable with this improved sensitivity are shown as pluses ($+$) in
Fig.~\ref{Easpects14ab} \cite{Easpects}.
\end{itemize}

\subsection{LEPII}
We discuss three supersymmetry signatures at LEPII: Higgs boson production,
chargino pair production, and selectron pair production.
\subsubsection{Lightest Higgs boson}
\begin{itemize}
\item The dominant production mechanism is: $e^+e^-\to Z^*\to Zh (h\to b\bar
b)$, where the Higgs boson decays into two $b$-jets and $b$-tagging is used
to reduce the light-jet background.
\item The cross section for the supersymmetric process is proportional to the
corresponding Standard Model cross section:\\
$\sigma_{susy}=\sin^2(\alpha-\beta)\sigma_{SM}\approx\sigma_{SM}$, where
the last result follows in models with radiative breaking, \ie, the lightest
supersymmetric Higgs boson looks a lot like the Standard Model Higgs boson.
\item The relevant branching fraction $B(h\to b\bar b)\approx B(H_{SM}\to b\bar
b)$, except when the $h\to\chi^0_1\chi^0_1$ channel is open (for $\lsim10\%$ of
the points) in which case the expected signal can be greatly eroded.
\item The cross section for Higgs boson production at LEPII for $\sqrt{s}=200$
and $210\GeV$ are shown in Fig.~\ref{higgs} \cite{Easpects}. The accumulation
of points corresponds to the Standard Model result; the points ``falling off"
the curves correspond to the opening of the $h\to\chi^0_1\chi^0_1$ channel.
\item The LEPII sensitivity is expected to be $\gsim(0.1-0.2)\pb$ (dashed lines
in Fig.~\ref{higgs}) \cite{Sopczak}, which implies a mass reach of
$m_h\lsim\sqrt{s}-95$.
\item We note that the Higgs-boson mass is the most directly useful piece of
information that could come out of LEPII. This is shown in
Fig.~\ref{Easpects19ab} \cite{Easpects}, where the Higgs-boson mass contours
are given and show that once $m_h$ is known, $\tan\beta$ would be determined
in terms of the chargino mass.
\end{itemize}

\subsubsection{Charginos}
\begin{itemize}
\item The production channel is: $e^+e^-\to \chi^+_1\chi^-_1\to 1l+2j$, where
the 1-lepton+2-jets signature (\ie, the ``mixed" signal) is used.
\item A problem with this channel occurs when $B(\chi^\pm_1\to l)\approx1$
and the chargino branching fraction into jets is strongly suppressed. This
phenomenon occurs in the $SU(5)\times U(1)$ models we consider.
\item The expected experimental sensitivity ($5\sigma$ signal over background)
is $(\sigma B)_{mixed}\gsim0.05\pb$ with ${\cal L}=500\ipb$ \cite{Grivaz}.
\item In $SU(5)\times U(1)$ supergravity this discovery channel has been first
studied in Ref.~\cite{LNPWZ}. The points in the still-allowed parameter space
which are reachable through this mode are denoted by crosses ($\times$) in
Fig.~\ref{Easpects14ab}.
\end{itemize}

\subsubsection{Sleptons}
\begin{itemize}
\item The production channel is: $e^+e^-\to\tilde e\tilde e\to ee
p\hskip-10pt/\hskip2pt$, where the lightest (right-handed) selectron is
the one dominantly produced.
\item The experimental sensitivity (at the $5\sigma$ level) is expected to be
$(\sigma B)_{dilepton}\gsim0.21\pb$ with ${\cal L}=500\ipb$
\cite{Dionisi,Easpects}. The reason for this large background is the
irreducible background cross section $\sigma(e^+e^-\to W^+W^-\to 2l)=0.9\pb$.
\item The points in the still-allowed parameter space which are reachable
through this mode are denoted by crosses ($\diamond$) in
Fig.~\ref{Easpects14ab} \cite{LNPWZ,Easpects}.
\item Note that selectrons consitute a deeper probe of the parameter space
than direct chargino searches (in the no-scale scenario). Similar remarks
apply to $\tilde\mu$ and $\tilde\tau$ pair production, although the cross
sections are somewhat smaller because of the loss of the $t$-channel diagrams.
(Sleptons are too heavy to be observable at LEPII in the dilaton scenario.)
\end{itemize}

\noindent $\bullet$ Note that if the Tevatron (1994--95) sees charginos,
then LEPII (1996) will be in business.

\subsection{DiTevatron}
\begin{itemize}
\item In the wake of the SSC demise, there has been a recent proposal to
upgrade the Tevatron \cite{vision,design,Amidei} as follows:
\begin{itemize}
\item $p\bar p$ collisions at $\sqrt{s}=4\TeV$
\item ${\cal L}=2\times10^{32}{\rm cm}^{-2}{\rm s}^{-1}
\to \int{\cal L}\sim 2\ifb/$year.
\item Would use single ring of SSC magnets, sized down to fit the Tevatron
tunnel. The Tevatron would be used for injection. The detectors would need
to be upgraded (as currently planed) but not replaced.
\item If this plan is approved, the new rings of magnets should be installed
when the Main Injector is put in place, and the machine could be doing physics
before the LHC turns on.
\end{itemize}
\item We have studied the possible reach of this machine for charginos via
the trilepton mode, squarks and gluinos via the missing energy signature,
and Higgs bosons produced in association with a $W$ or $Z$ boson
\cite{DiTevatron}. With an integrated luminosity of $5\ifb$, the reach for
chargino masses is expected to be $210\,(150)\GeV$ in the moduli (dilaton)
scenario. This is exemplified in Figs.~\ref{Figure2},\ref{Figure3} where the
corresponding reach at the Tevatron is also shown. The corresponding reach
in squark and gluino masses is estimated to be $\sim700\GeV$ and is depicted
in Fig.~\ref{Figure5} in terms of the significance for such a signal. The
lightest Higgs boson could also be searched up to a mass of $\sim120\GeV$
\cite{GH,DiTevatron}.
\item All in all, the doubling of the Tevatron energy to the DiTevatron should
allow one to probe a large fraction of parameter space of $SU(5)\times U(1)$
supergravity.
\end{itemize}

\section{Conclusions}
\begin{itemize}
\item True string phenomenology is very powerful. Once a vacuum is singled
out, every parameter of the model is in principle calculable.
\item We have discussed a two-parameter (plus $m_t$), very predictive,
string-inspired $SU(5)\times U(1)$ supergravity model.
\item With the string-inspired assumptions for the supersymmetry breaking
scenarios, several experimental tests have been worked out in detail.
\item Experimental outlook
\begin{description}
\item 1994:\\
Tevatron (trileptons)\\
CLEO ($b\to s\gamma$)\\
LEPI ($m_h$)
\item 1996:\\
 Brookhaven $(g-2)_\mu$\\
 LEPII (Higgs, charginos, sleptons)\\
 MACRO (neutralinos)
\item 2000(?):\\
DiTevatron (charginos, squarks, gluinos, Higgs)
\item 2005(?):\\
LHC (Higgs, squarks, gluinos, charginos)
\end{description}
\item We should remark that the ``real" string model, which incorporates all
the features we would like to have in a supergravity model, is yet to be built.
\item Realistic supersymmetry breaking scenarios derived from string remain as
the largest stumbling block to a ``Theory of Everything".
\end{itemize}

\newpage

\section*{Figure Captions}
\begin{enumerate}
\item The external string states are mapped onto the world-sheet by a
conformal transformation. The crosses represent the vertex operators which
describe the mapped string states.
\label{vertex}
\item String perturbation theory is an expansion in the topology of the
two-dimensional world-sheet. Higher orders in perturbation theory are
represented by surfaces with increasingly larger number of handles.
Note that at each order in perturbation theory there is only one string
diagram.
\label{topology}
\item Dimension five operator mediating proton decay in $SU(5)\times U(1)$.
This diagram is suppressed because of the lack of $H_3$, $\bar H_3$ mixing.
\label{d5pd}
\item  The top-quark Yukawa coupling at the unification scale versus $m_t$ for
fixed values of $\tan\beta$. The horizontal lines indicate a possible
range of string predictions for the top-quark Yukawa coupling.
\label{yuks}
\item The running of the gauge couplings in $SU(5)\times U(1)$ supergravity for
$\alpha_3(M_Z)=0.118$ (solid lines). The intermediate-scale particle masses
have been derived using the gauge coupling RGEs to achieve unification at
$M_U=10^{18}\GeV$. The case with no intermediate-scale particles (dotted lines)
is also shown; here $M_U\approx10^{16}\GeV$.
\label{proc_erice1}
\item A ``flow chart" of the various steps typically followed in determining
the theoretically allowed parameter space in a supergravity model. The input
parameters are $m_t,\tan\beta,m_{1/2},m_0,A$.
\label{chart}
\item Typical variation of the tree-level scalar potential as the
renormalization scale $Q$ is decreased. One starts with no minimum for
$Q>Q_0$, and ends up with runaway minima for $Q<Q_1$.
\label{potential}
\item Schematic variation of the minima of the tree-level Higgs potential
as the renormalization scale $Q$ is lowered.
\label{vevs}
\item Running of the scalar squared masses in supergravity for a typical
choice of model parameters (indicated). Note that the electroweak symmetry
is broken radiatively, and the squark and slepton squared masses remain
positive.
\label{runnings}
\item
The parameter space for no-scale $SU(5)\times U(1)$ supergravity (moduli
scenario) in the $(m_{\chi^\pm_1},\tan\beta)$ plane for $m_t=150,170\GeV$. The
periods indicate points that passed all theoretical and experimental
constraints, the pluses fail the $b\to s\gamma$ constraint, the crosses fail
the $(g-2)_\mu$ constraint, the diamonds fail the neutrino telescopes (NT)
constraint, the squares fail the $\epsilon_1-\epsilon_b$ constraint, and the
octagons fail the updated Higgs-boson mass constraint. The reference dashed
line highlights $m_{\chi^\pm_1}=100\GeV$, which is the direct reach of LEPII
for chargino masses. Note that when various symbols overlap a more complex
symbol is obtained.
\label{moduliPS}
\item
The parameter space for no-scale $SU(5)\times U(1)$ supergravity (dilaton
scenario) in the $(m_{\chi^\pm_1},\tan\beta)$ plane for $m_t=150,170\GeV$. The
periods indicate points that passed all theoretical and experimental
constraints, the pluses fail the $b\to s\gamma$ constraint, the crosses fail
the $(g-2)_\mu$ constraint, the diamonds fail the neutrino telescopes (NT)
constraint, the squares fail the $\epsilon_1-\epsilon_b$ constraint, and the
octagons fail the updated Higgs-boson mass constraint. The reference dashed
line highlights $m_{\chi^\pm_1}=100\GeV$, which is the direct reach of LEPII
for chargino masses. Note that when various symbols overlap a more complex
symbol is obtained.
\label{dilatonPS}
\item The largest one-loop contributions to the $b\to s\gamma$ process in
supersymmetric models. The last diagram can interfere destructively with
the first two for $\mu>0$.
\label{bsgd}
\item The calculated values of $B(b\to s\gamma)$ in the moduli scenario.
Note the destructive interference effect for $\mu>0$. The present CLEOII
limits are as indicated.
\label{bsgr}
\item The one-loop supersymmetric contributions to the anomalous magnetic
moment of the muon.
\label{g-2d}
\item The calculated values of $(g-2)_\mu$ in the moduli scenario. Note that
large enhancements occur for not too small values of $\tan\beta$.
\label{g-2r}
\item Main diagrams contributing to the parameter $\epsilon_1$ in
supersymmetric models.
\label{eps1}
\item Main diagrams contributing to the parameter $\epsilon_b$ in
supersymmetric models.
\label{epsb}
\item An example of the effect of a light chargino on the electroweak parameter
$\epsilon_1$. A light chargino shifts $\epsilon_1$ negatively and thus
compensates for a large top-quark mass.
\label{eps1inset}
\item The calculated values of $\epsilon_1$ and $\epsilon_b$ in the
$SU(5)\times U(1)$ model with moduli scenario. The various experimental
ellipses indicate a preference for lighter top-quark masses. The size of the
ellipses is expected to be reduced by a factor of two with the 93+94 LEP data.
\label{Easpects5a}
\item  The predicted upwardly-moving muon flux for Earth capture of galactic
halo neutralinos, and the present Kamiokande flux upper limit. One can see the
${\rm Fe}^{56}$ capture enhancement, and that the data already impose some
(although small) restrictions on parameter space. An improvement in sensitivity
by a factor of 2--10 is expected with MACRO.
\label{NT}
\item The rate for trilepton events versus the chargino mass in the
$SU(5)\times U(1)$ ``moduli scenario" (the results are somewhat smaller in the
``dilaton scenario").
\label{trilep}
\item The remaining allowed parameter space in $SU(5)\times U(1)$ supergravity
in the moduli (top plots) and dilaton (bottom plots) scenarios. Points
accesible by trilepton searches are denoted by pluses ($+$), whereas those
accessible by chargino and selectron searches at LEPII are donoted by crosses
($\times$) and diamonds ($\diamond$), respectively. The contours are of the
lightest Higgs boson mass.
\label{Easpects14ab}
\item
The cross section for Higgs boson production at LEPII for $\sqrt{s}=200$
and $210\GeV$. The accumulation of points corresponds to the Standard Model
result; the points ``falling off" the curves correspond to the opening of the
$h\to\chi^0_1\chi^0_1$ channel.
\label{higgs}
\item The Higgs-boson mass contours in the $(m_{\chi^\pm_1},\tan\beta)$ plane
for both scenarios in $SU(5)\times U(1)$ supergravity. Once $m_h$ is known,
$\tan\beta$ would be determined as a function of the chargino mass. The
Higgs-boson mass would be the most useful piece of information to come out
of LEPII.
\label{Easpects19ab}
\item Trilepton yield ($\sigma\times B$) versus chargino mass in chargino
production in $p\bar p$ collisions. The dots define the range of parameters
allowed within the string-inspired $SU(5)\times U(1)$ supergravity model
(for $m^{\rm pole}_t=157\GeV$) with moduli scenario for supersymmetry breaking.
Results are shown for each sign of the Higgs mixing parameter $\mu$. The upper
(lower) plots show the limits which could be reached at the Tevatron
(DiTevatron). The sensitivity limits are for the indicated integrated
luminosities .
\label{Figure2}
\item Trilepton yield ($\sigma\times B$) versus chargino mass in chargino
production in $p\bar p$ collisions. The dots define the range of parameters
allowed within the string-inspired $SU(5)\times U(1)$ supergravity model
(for $m^{\rm pole}_t=157\GeV$) with dilaton scenario for supersymmetry
breaking. Results are shown for each sign of the Higgs mixing parameter $\mu$.
The upper (lower) plots show the limits which could be reached at the Tevatron
(DiTevatron). The sensitivity limits are for the indicated integrated
luminosities.
\label{Figure3}
\item Statistical significance for gluino and squark events at the DiTevatron
with ${\cal L}=5\ifb$. (The significance varies with ${\cal L}$ as $\sqrt{\cal
L}$.) These eventes were selected by the criteria $p_\perp>150\,{\rm GeV}$, and
4 jets with $p_\perp>40\,{\rm GeV}$.
Bands are shown for signal $S$ from gluino pairs, and squark/gluino
combinations for parameters which are consistent with the minimal $SU(5)$
supergravity model and the $SU(5)\times U(1)$ supergravity models,
respectively. The background $B$ is calculated from $Z\to\nu\bar\nu$;
the bands provide for a factor of 5 deterioration of $S/B$ ratio
due to additional backgrounds or inefficiencies.
\label{Figure5}
\end{enumerate}

\newpage
\normalsize
\begin{thebibliography}{99}
\bibitem{GSW} M. B. Green, J. H. Schwarz, and E. Witten, {\em Superstring
Theory}, Vol. 1 (Cambridge University Press, Cambridge 1987).
\bibitem{BK} Z. Bern and D. Kosower, \PRL{66}{91}{1669}, \NPB{362}{91}{389},
\NPB{379}{92}{451}.
\bibitem{FFF} I. Antoniadis, C. Bachas, and C. Kounnas, Nucl. Phys. B
{\bf 289} (1987) 87; I. Antoniadis and C. Bachas, Nucl. Phys. B {\bf298} (1988)
586; H. Kawai, D.C. Lewellen, and S.H.-H. Tye, Phys. Rev. Lett. {\bf57} (1986)
1832; Phys. Rev. D {\bf34} (1986) 3794; Nucl. Phys. B {\bf288} (1987) 1;
R. Bluhm, L. Dolan, and P. Goddard, Nucl. Phys. B {\bf309} (1988) 330;
H. Dreiner, J. L. Lopez, D. V. Nanopoulos, and D. Reiss, Nucl. Phys. B
{\bf 320} (1989) 401.
\bibitem{Kaplunovsky} V. Kaplunovsky, \NPB{307}{88}{145}.
\bibitem{thresholds} For explicit examples see, I. Antoniadis, J. Ellis, R.
Lacaze, and \DVN, \PLB{268}{91}{188}; S. Kalara, \JL, and \DVN,
\PLB{269}{91}{84}.
\bibitem{KLN} S. Kalara, J. Lopez and D.V. Nanopoulos, Nucl. Phys. B
{\bf 353} (1991) 650.
\bibitem{decisive} \JL\ and \DVN, \PLB{251}{90}{73} and \PLB{268}{91}{359}.
\bibitem{GO}For a review see \eg, P. Goddard and D. Olive, \IJMP{1}{86}{303}.
\bibitem{Slansky} R. Slansky, \PRT{79}{81}{1}. For a comprehensive listing
see W. McKay and J. Patera, {\em Tables of dimensions, indices, and branching
rules for representations of simple algebras} (Dekker, New York, 1981).
\bibitem{ELN} J. Ellis, J. Lopez, and \DVN, \PLB{245}{90}{375}; A. Font, L.
Ib\'a\~nez, and F. Quevedo, \NPB{345}{90}{389}.
\bibitem{BB} K. S. Babu and S. Barr, BA-94-04 (1994).
\bibitem{Lewellen} D. Lewellen, \NPB{337}{90}{61}; J. A. Schwarz,
\PRD{42}{90}{1777}.
\bibitem{revitalized} I. Antoniadis, J. Ellis, J. Hagelin, and \DVN,
\PLB{194}{87}{231}.
\bibitem{ALR} I. Antoniadis, G. Leontaris and J. Rizos, Phys. Lett. B
{\bf 245} (1990) 161.
\bibitem{FNY}  A. Faraggi, D.V. Nanopoulos and K. Yuan, Nucl. Phys.
B {\bf 335} (1990) 347.
\bibitem{Alon} A. Faraggi, \PLB{278}{92}{131}, \PLB{274}{92}{47},
\NPB{387}{92}{239}.
\bibitem{MPM} A. Masiero, \DVN, K. Tamvakis, and T. Yanagida,
\PLB{115}{82}{380}; B. Grinstein, \NPB{206}{82}{387}.
\bibitem{chorus+ENO+ELNO} J. Ellis, \JL, and \DVN, \PLB{292}{92}{189};
J. Ellis, \DVN, and K. Olive, \PLB{300}{93}{121};
J. Ellis, \JL, \DVN, and K. Olive, \PLB{308}{93}{70}.
\bibitem{faspects} J. Ellis, J. Hagelin, S. Kelley, and \DVN,
\NPB{311}{88/89}{1}.
\bibitem{revamped} I. Antoniadis, J. Ellis, J. Hagelin, and \DVN,
\PLB{231}{89}{65}.
\bibitem{search} J.L Lopez, D.V. Nanopoulos and K. Yuan, Nucl. Phys. B
{\bf 399} (1993) 654.
\bibitem{Martin} S. P. Martin, Phys. Rev. D {\bf 46} (1992) 2769.
\bibitem{LNZI}\JL, \DVN, and A. Zichichi, \PRD{49}{94}{343}.
\bibitem{EVA}S. Kelley, \JL, and \DVN, \PLB{261}{91}{424}.
\bibitem{t-paper} \JL, \DVN, and A. Zichichi, \TAMU{78/93} (to appear in
Phys. Lett. B).
\bibitem{susyx} L. Dixon, in Proceedings of The Rice Meeting, ed. by B.
Bonner and H. Miettinen (World Scientific, 1990), p. 811, and references
therein.
\bibitem{Ross} A. de la Macorra and G. G. Ross, \NPB{404}{93}{321} and
\PLB{325}{94}{85}.
\bibitem{Casas} See \eg, J. Casas, Z. Lalak, C. Mu\~noz, and G. Ross,
\NPB{347}{90}{243}; L. Ib\'a\~nez and D. L\"ust, \NPB{382}{92}{305};
B. de Carlos, J. Casas, and C. Mu\~noz, \NPB{399}{93}{623}
and \PLB{299}{93}{234}.
\bibitem{Ig} I. Antoniadis, C. Bachas, D. Lewellen, and T. Tomaras,
\PLB{207}{88}{441}; I. Antoniadis, \PLB{246}{90}{377}; I. Antoniadis, C.
Mu\~noz, and M. Quiros, \NPB{397}{93}{515}.
\bibitem{KL}V. Kaplunovsky and J. Louis, \PLB{306}{93}{269}.
\bibitem{BIM} A. Brignole, L. Ib\'a\~nez, and C. Mu\~noz, FTUAM-26/93 (August
1993).
\bibitem{EN}J. Ellis and \DVN, \PLB{110}{82}{44}.
\bibitem{LNY94} \JL, \DVN, and K. Yuan, \TAMU{14/94}.
\bibitem{EWx}L. Ib\'a\~nez and G. Ross, \PLB{110}{82}{215}; K. Inoue, \etal,
Prog. Theor. Phys. 68 (1982) 927; L. Ib\'a\~nez, \NPB{218}{83}{514} and
\PLB{118}{82}{73}; H. P. Nilles, \NPB{217}{83}{366}; J. Ellis, \DVN, and
K. Tamvakis, \PLB{121}{83}{123}; J. Ellis, J. Hagelin, \DVN, and K. Tamvakis,
\PLB{125}{83}{275}; L. Alvarez-Gaum\'e, J. Polchinski, and M. Wise,
\NPB{221}{83}{495}; L. Iba\~n\'ez and C. L\'opez, \PLB{126}{83}{54} and
\NPB{233}{84}{545}; C. Kounnas, A. Lahanas, \DVN, and M. Quir\'os,
\PLB{132}{83}{95} and C. Kounnas, A. Lahanas, \DVN, and M. Quir\'os,
\NPB{236}{84}{438}.
\bibitem{LN}For a review see A. B. Lahanas and D. V. Nanopoulos,
\PRT{145}{87}{1}.
\bibitem{preLEP} G. Gamberini, G. Ridolfi, and F. Zwirner, \NPB{331}{90}{331};
J. Ellis and F. Zwirner, \NPB{338}{90}{317}.
\bibitem{LEPera}S. Kelley, \JL, \DVN, H. Pois, and K. Yuan, \PLB{273}{91}{423};
G. Ross and R. Roberts, \NPB{377}{92}{571}; M. Drees and M.M. Nojiri,
\NPB{369}{92}{54}; K. Inoue, M. Kawasaki, M. Yamaguchi, and T. Yanagida,
\PRD{45}{92}{328}; R. Arnowitt and P. Nath, \PRL{69}{92}{725}; P. Nath and
R. Arnowitt, \PLB{287}{92}{89}.
\bibitem{aspects}S. Kelley, \JL, \DVN, H. Pois, and K. Yuan, \NPB{398}{93}{3}.
\bibitem{recent} R. Roberts and L. Roszkowski, \PLB{309}{93}{329};
M. Olechowski and S. Pokorski, \NPB{404}{93}{590}; B. de Carlos and J. Casas,
\PLB{309}{93}{320}; M. Carena, L. Clavelli, D. Matalliotakis, H. Nilles,
and C. Wagner, \PLB{317}{93}{346}; G. Leontaris, \PLB{317}{93}{569};
S. Martin and P. Ramond, \PRD{48}{93}{5365}; D. Casta\~no, E.
Piard,  and P. Ramond, UFIFT-HP-93-18 (August 1993); W. de Boer, R.
Ehret,  and D. Kazakov, IEKP-KA/93-13 (August 1993); A. Faraggi and B.
Grinstein, SSCL-Preprint-496 (August 1993); M. Bastero-Gil, V. Manias, and J.
Perez-Mercader, LAEFF-93/012 (September 1993); M. Carena, M. Olechowski, S.
Pokorski, and C. Wagner, CERN-TH.7060/93 (October 1993); V. Barger, M. Berger,
and P. Ohmann, MAD/PH/801 (November 1993); A. Lahanas, K. Tamvakis, and N.
Tracas, CERN-TH.7089/93 (November 1993); G. Kane, C. Kolda, L. Roszkowski, and
J. Wells, UM-TH-93-24 (December 1993).
\bibitem{Layman}For an elementary introduction and a review of this field see
\JL, \DVN, and \AZ, Rivista del Nuovo Cimento, {\bf17} (1994) 1.
\bibitem{LNZII}\JL, \DVN, and A. Zichichi, \PLB{319}{93}{451}.
\bibitem{BLM} R. Barbieri, J. Louis, and M. Moretti, \PLB{312}{93}{451} and
\PLB{316}{93}{632}(E).
\bibitem{D0} S. Abachi, \etal (D0 Collaboration), \PRL{72}{94}{2138}.
\bibitem{Altarelli} G. Altarelli, in {\em Proceedings of the International
Europhysics Conference on High Energy Physics}, Marseille, France, July 22--28,
1993, edited by J. Carr and M. Perrottet (Editions Frontieres, Gif-sur-Yvette,
1993) CERN-TH.7045/93 (October 1993), G. Altarelli, private communication; J.
Ellis, G. L. Fogli, and E. Lisi, CERN-TH.7116/93 (December 1993); J. Ellis,
private communication; P. Langacker and N. Polonsky, UPR-0594T (February 1994);
V. Novikov, L. Okun, A. Rozanov, and M. Vysotsky, CERN-TH.7217/94 (April 1994).
\bibitem{CDF} The CDF Collaboration, ``Evidence for top quark production in
$\bar p p$ collisions at $\sqrt{s} = 1.8\TeV$, Fermilab-Pub-94/097-E (April
1994).
\bibitem{Easpects} \JL, \DVN, G. Park, X. Wang, and A. Zichichi, \TAMU{74/93}.
\bibitem{bsgamma+bsg-eps}\JL, \DVN, and G.~T.~Park, \PRD{48}{93}{R974};
\JL, \DVN, G.~T.~Park, and A. Zichichi, \PRD{49}{94}{355}.
\bibitem{Thorndike} E. Thorndike, Bull. Am. Phys. Soc. {\bf38}, 922 (1993);
R. Ammar, \etal, CLEO Collaboration, \PRL{71}{93}{674}.
\bibitem{g-2}\JL, \DVN, and X. Wang, \PRD{49}{94}{366}.
\bibitem{oldg}J. Bailey \etal, \NPB{150}{79}{1}.
\bibitem{kinoII} For a recent review see, T. Kinoshita, Z. Phys. C{\bf56}
(1992) S80, and in {\em Frontiers of High Energy Spin Physics}, Proceedings
of the 10th International Symposium on High Energy Spin Physics, edited by
T. Hasegawa, N. Horikawa, A. Masaike, and S. Sawada (Universal Academy Press,
1993).
\bibitem{newg}M. May, in AIP Conf. Proc. USA Vol. 176 (AIP, New York, 1988)
p. 1168; B. L. Roberts, Z. Phys. {\bf C56} (1992) S101.
\bibitem{ABC}G. Altarelli and R. Barbieri, \PLB{253}{90}{161}; G. Altarelli, R.
Barbieri, and S. Jadach, \NPB{369}{92}{3}; R. Barbieri, M. Frigeni, and F.
Caravaglios, \PLB{279}{92}{169}; G. Altarelli, R. Barbieri, and F. Caravaglios,
\NPB{405}{93}{3} and \PLB{314}{93}{357}; \JL, \DVN, G. Park, H. Pois, and K.
Yuan, \PRD{48}{93}{3297}; \JL, \DVN, G.~T.~Park, and A. Zichichi,
\PRD{49}{94}{4835}.
\bibitem{NT}R. Gandhi, \JL, \DVN, K. Yuan, and A. Zichichi, \PRD{49}{94}{3691}.
\bibitem{LNWZ}\JL, \DVN, X. Wang, and A. Zichichi, \PRD{48}{93}{2062}.
\bibitem{trileptons} J. Ellis, J. Hagelin, \DVN, and M. Srednicki,
\PLB{127}{83}{233}; H. Baer and X. Tata, \PLB{155}{85}{278}; H. Baer, K.
Hagiwara, and X. Tata, \PRL{57}{86}{294}, \PRD{35}{87}{1598}; P. Nath and R.
Arnowitt, \MODA{2}{87}{331}; R. Barbieri, F. Caravaglios, M. Frigeni, and M.
Mangano, \NPB{367}{91}{28}; H. Baer and X. Tata, \PRD{47}{93}{2739}; H. Baer,
C. Kao, and X. Tata, \PRD{48}{93}{5175}.
\bibitem{LNWZ}\JL, \DVN, X. Wang, and A. Zichichi, \PRD{48}{93}{2062}.
\bibitem{Kato} Talk given by Y. Kato (CDF Collaboration) at the 9th Topical
Workshop on Proton-Antiproton Collider Physics, Tsukuba, Japan, October 1993.
\bibitem{Sopczak} A. Sopczak, L3 note 1543 (November 1993).
\bibitem{Grivaz}J.-F. Grivaz, LAL preprint 92-64 (November 1992).
\bibitem{LNPWZ}\JL, \DVN, H. Pois, X. Wang, and A. Zichichi,
\PRD{48}{93}{4062}.
\bibitem{Dionisi}C. Dionisi, \etal, in Proceedings of the ECFA Workshop on LEP
200, Aachen, 1986, ed. by A. B\"ohm and W. Hoogland, p. 380.
\bibitem{vision} ``A vision for high energy physics", T. Kamon, J. L. Lopez, P.
McIntyre, and J. White, CTP-TAMU-11/94 (February 1994).
\bibitem{design} ``Conceptual design for a Tevatron upgrade to 2 TeV beams and
luminosity $>10^{33}{\rm cm}^{-2}{\rm s}^{-1}$", G. Jackson, J. Strait,
D. Amidei, G. W. Foster, D. Baden, S. Holmes, D. Finley, and J. Theilacker
(March 1994).
\bibitem{Amidei} ``Top factory at the Tevatron", D. Amidei, \etal
(March 1994).
\bibitem{DiTevatron} ``Supersymmetry at the DiTevatron", T. Kamon, etal,
\TAMU{19/94}.
\bibitem{GH} J.F. Gunion and T. Han, UCD-94-10 (April 1994); A. Stange, W.
Marciano, and S. Willenbrock, ILL-TH-94-8 (April 1994).
\end{thebibliography}
\end{document}